\documentclass[letterpaper]{article} 
\usepackage{aaai2026}  
\usepackage{times}  
\usepackage{helvet}  
\usepackage{courier}  
\usepackage[hyphens]{url}  
\usepackage{graphicx} 
\usepackage{natbib}  
\usepackage{caption} 
\usepackage{algorithm}
\usepackage{algorithmic}
\usepackage{booktabs}
\usepackage{amsmath}
\usepackage{placeins}

\usepackage{algorithm}
\usepackage{algorithmic}
\usepackage{float}

\usepackage{xcolor}
\newcommand{\answerYes}[1]{\textcolor{blue}{#1}} 
\newcommand{\answerNo}[1]{\textcolor{teal}{#1}} 
\newcommand{\answerNA}[1]{\textcolor{gray}{#1}}

\usepackage{newfloat}
\usepackage{listings}
\DeclareCaptionStyle{ruled}{labelfont=normalfont,labelsep=colon,strut=off} 
\floatstyle{ruled}
\newfloat{listing}{tb}{lst}{}
\floatname{listing}{Listing}
\title{WhichTok? Comparing Three TikTok Data Acquisition Tools}
\author{
    Gayoung Jeon\textsuperscript{\rm 1}\protect\thanks{Corresponding author.},
    Cameron Moy\textsuperscript{\rm 1},
    Silvia Téliz\textsuperscript{\rm 1},
    Cristina Monzer\textsuperscript{\rm 1},
    Nicolette Alayón\textsuperscript{\rm 2},
    Deen Freelon\textsuperscript{\rm 1}
}

\affiliations{
    \textsuperscript{\rm 1}University of Pennsylvania\\
    \{gjeon, moycam, silteliz, cmonzer, dfreelon\}@upenn.edu\\
    \textsuperscript{\rm 2}Northwestern University\\
    nicolettealayon2023@u.northwestern.edu
}

\begin{document}

\maketitle

\begin{abstract}
TikTok's global growth has made it a prime platform for both entertainment and political discourse, prompting increased social science research. However, this rapidly evolving research field faces a fundamental reproducibility crisis. TikTok's opaque algorithmic systems hinder researchers from drawing meaningful empirical inferences, while the lack of standardized data collection methods compounds these challenges. This study addresses these methodological gaps by systematically comparing three data collection tools ---the official TikTok Research API, Pyktok, and Apify. We evaluated five endpoints, \textit{User, Hashtag, Keyword, Comment, and Related Video}. Results show substantial cross-tool differences, especially for hashtag and keyword searches. The Research API uses back-end API calls, whereas Apify and Pyktok rely on front-end web scraping, producing systematic differences in the time periods and popularity levels represented in retrieved content. The three tools yielded  comprehensive and consistent results only for the user endpoint. Our results question whether these tools can acquire truly \textit{random[-ized]} samples, as they introduce methodological confounds that may compromise research validity in ways not yet fully understood. Based on these results, we offer methodological, transparency, and ethical recommendations and guidelines to increase TikTok research quality.
\end{abstract}


\section{Introduction}

TikTok has gained popularity as a global video and social media platform, primarily due to its content recommendation algorithm and infinite scroll feature. The platform was the second most used social media app (second only to YouTube) among teens in 2024 \cite{favero_teens_2024, tidy_tiktok_2020}. Currently, TikTok's popularity has garnered significant academic attention at the ICWSM, spanning topics including global political issues, misinformation, hate speech, and algorithmic bias \cite{ji_revealing_2025, luik_media_2025, rejeb_mapping_2024, sharevski2025lowvision, galdeman2025climate, ungless2025censorship, yang2025addiction, pinto_fighting_2024}. Yet, as research interest in TikTok data continues to accelerate, we still know little about the quality of data collected from it. The empirical reality researchers currently navigate is that APIs represent just one data access alternative among many \cite{Freelon2025}. A critical concern in TikTok research is whether different data collection tools retrieve consistent data for the same targets of interest, e.g. specific events, hashtags, or individual accounts. Without rigorous answers to this question, TikTok research findings risk reflecting biases from methodological choices rather than accurately representing the phenomena under study. 

To address this, we systematically test three TikTok data collection methods that span the main access pathways available for data collection at scale: the official TikTok Research API, Pyktok (an open-source TikTok data collection package for Python), and Apify (a commercial data collection platform). After collecting data for a month across five different data collection endpoints (hashtags, keyword search, user pages, related videos, and comments), we find that each API produces dramatically different datasets even when all conditions (API parameters) are held constant. These results confirm that TikTok research faces reproducibility and validity issues caused by algorithmically influenced sampling and the under-documentation of our data collection tools' limitations. 

The empirical results of this study coincide with our review of existing TikTok studies. Researchers have leveraged a wide variety of data collection methods and followed substantially different standards when releasing their data collection protocols \cite{bickham2026tied, goanta2026great, xu2026anti}. While some researchers go to great lengths to explain their sampling methods, others omit these details altogether. The lack of consistency and transparency among TikTok data collection methods within web and social science research has led to what we argue are serious reproducibility and internal and external validity issues. Thus, in addition to our empirical contributions, we offer practical recommendations for TikTok data collection as well as transparency and ethical guidance.

\section{Related Work}

We conducted a search for peer-reviewed articles in the social sciences that used TikTok as their main data source\footnote{The selection protocol for relevant publications consisted of searching for ‘TikTok’ in the abstract, title, or keywords, and ‘quantitative’ in any of the article sections in EBSCO’s Communication Source, Academic Search Premier, and Political Science Complete databases; and Scopus (in which results were limited to Social Sciences). The results were limited to full academic peer-reviewed articles published between 2020 and 2025. This search yielded a total of 239 publications, out of which only 90 used TikTok content as their main data source}. We found that researchers across various disciplines, from business and psychology to communication and human-computer interaction, use TikTok as a window into the social world. Studies have analyzed the platform's content to explore users' attitudes and behaviors \cite{ji_revealing_2025, luik_media_2025, steinke_women_2024}, political actors' usage of the platform \cite{dominguez-garcia_emotion_2025, caro-castano_self-presentation_2025, cartes-barroso_attracting_2025}, and the communication of health-related issues \cite{lee_identification_2025, sun_vaping_2023}, among other topics. Content analysis is a common approach for processing TikTok data \cite{kanthawala_its_2022, rejeb_mapping_2024}.

Algorithmic audit studies have shed some light on TikTok's curation criteria. One study found that a user's location, language, liking behavior, and followed accounts are personalization factors that influence what users see in their FYP (For You Page) \cite{boeker_empirical_2022}. Furthermore, \citet{mousavi_auditing_2024} reports that TikTok provides eight different explanations for why a video shows up in the FYP, which appear as pop-ups when browsing the app. These possible explanations are content-based (the user shared/commented/liked/watched similar videos or follows a specific account) and engagement-based (the video was recently posted, is popular in the user's country, and has a length that the user tends to like). However, \citet{mousavi_auditing_2024} argue that these explanations are often unrelated to the user's actual behavior, rendering the platform's attempt to increase its algorithm's transparency inadequate. Without a comprehensive understanding of TikTok's content recommendation algorithm, sampling processes remain uncertain.

In spite of TikTok's opaque recommendation system, researchers have collected data through three main channels: manual collection through the platform's user interface (UI) (i.e., the search page or FYP), third-party web scrapers, or application programming interfaces (APIs). Given the heavy influence of algorithmic recommendations on TikTok's content feeds, it is difficult to discern whether any of these approaches has the capacity to yield samples of videos from which one may draw generalizable conclusions. In this section, we identify the strengths and limitations of each of these three approaches and discuss their susceptibility to algorithmic bias.

\subsection{Manual UI-based data collection}

Manual UI-based data collection methods consist of downloading video, text, and other metadata that appear on TikTok's search page or FYP feed. Three main approaches qualify here, the first of which obtains data through the platform's search page through keyword, hashtag, or user searches. Studies that follow this first approach, search for relevant keywords---for instance, ‘medical education’ and ‘medical review’---and select the \textit{n} most-viewed videos as their sample \cite{izquierdo-condoy_assessing_2025}. The second approach samples from the FYP by scrolling through the page under conditions simulating use by end-users. Such studies have attempted to mitigate the algorithm's influence on the app's search results through the usage of \textit{sockpuppet} accounts. This strategy consists of creating a TikTok account that impersonates a user with specific characteristics. For instance, to explore the self-expression of pregnant minors, scholars created a new TikTok account using the demographic information of a fictional 14-year-old girl \cite{suarez-alvarez_communicational_2022}. This account was used to search for hashtags related to adolescent pregnancy and identify popular accounts that posted about the topic. Kaplan et al. used sockpuppet accounts to simulate users' control over their For You Page. They found that swiping out or reporting content as not interests reduced recommendations but that viewing could subsequently restore topical personalization \cite{kaplan2026foryou}.

Collecting data from TikTok's search page manually offers a relatively low learning curve, as it does not require computer programming skills. However, this convenience comes with a critical limitation: it remains unclear how user personalization factors influence the search page's results. Supporting this concern, recent surveys show that users themselves recognize the algorithmic influence on their feeds---and ironically, many report that this very personalized content is why they prefer TikTok over traditional search engines \cite{adobe_express_using_2024, diep_tiktok_2025, kaplan2026foryou}. Although TikTok auditing studies provide valuable insights, their focus on the \#FYP leaves unanswered the question of how the hyper-personalized algorithm affects search page results. This uncertainty surrounding the impact of personalized algorithms on data collection has raised important questions about the validity of research findings.

\subsection{Scraping-based data collection}

Web scraping techniques use the same data endpoints as manual UI data collection but differ by automating the process. Common web scrapers used by the studies in our literature review are \textit{Apify} \cite{infante_pineda_journalists_2025, meza_idols_2023}, \textit{Pyktok} \cite{scharlach_how_2024}, \textit{Octoparse} \cite{jerin_mental_2024}, and \textit{Fanpage Karma} \cite{londono-moreno_peruvian_2025}. Fanpage Karma and Octoparse provide profile- and page-level data and do not require programming skills. However, as these tools were designed for social media management, their costs are high for collecting large datasets \cite{fanpagekarma_professional_, octoparse_web_}. In contrast, Apify and Pyktok are more accessible for researchers in terms of cost, programming requirements, and flexibility. We will focus on evaluating the strengths and limitations of these two tools, given their suitability for research purposes.

Apify is a cloud-based platform offering off-the-shelf programs (known as \textit{actors}) that extract information from popular websites, including TikTok \cite{apify_apify_2025}. As Apify supports data collection through both a web interface and APIs, it is accessible to those with varying levels of computational skills and offers relatively low costs. \citet{meza_idols_2023} employ Apify to extract over 22,000 TikTok videos for a content analysis focusing on the communication strategies of TikTok celebrities. \citet{infante_pineda_journalists_2025}  used it to extract 657 videos for a content analysis on the communication style of journalists on TikTok. Nevertheless, the tool provides limited transparency regarding the data returned. Many Apify scrapers lack clear disclosure of how they retrieve TikTok data. Some rely on scraping web pages via HTTP requests of TikTok URLs (i.e., from the platform's front-end pages), others extract data via undocumented TikTok API endpoints, and some combine both approaches \cite{apify_web_2025}. Furthermore, the scrapers rely on TikTok's web interface, making it vulnerable to platform updates. Such lack of transparency about data access methods may put researchers at risk of obtaining inconsistent results \cite{pearson_beyond_2025}.

Unlike commercial scraping services, Pyktok provides an open-source and cost-free data collection solution geared for academic researchers. Pyktok is a Python library that ``pulls data directly from the JSON objects embedded in TikTok pages and from hidden APIs with no public documentation'' \cite{freelon_pyktok_nodate}. Collecting data with this tool requires familiarity with the Python programming language, a limiting factor for researchers who do not have coding experience. Perhaps because of this limitation, we found only one paper that uses this tool. This qualitative study explored depoliticization strategies on social media platforms and collected metadata for 360 videos using Pyktok \cite{scharlach_how_2024}. Since Pyktok is open-source, its data retrieval methods are fully visible in its repository, an advantage over the opaque Apify scrapers. However, similarly to Apify, the tool's reliance on TikTok's web interface makes it vulnerable to platform updates, and its continued functionality depends on community maintenance. Nevertheless, by giving researchers greater control over their collection environment and avoiding platform-specific subscription fees, Pyktok represents a flexible and sustainable option for TikTok data collection.

\subsection{API-based data collection}

An application programming interface (API) is a system that allows users to programmatically access platform data. In 2023, TikTok launched its Research API, which provides account and content-level data \cite{tiktok_research_2025}. The Research API requires programming expertise to design queries and process responses, and its availability is restricted to researchers based in the United States and Europe that receive approval. These constraints make the Research API less accessible than manual UI-based data collection methods.

\citet{zhang_impact_2025} used TikTok's Research API to collect 876 videos from hashtag queries \cite{zhang_impact_2025}, while \citet{santaolalla-rueda_potaxies_2024} used the API in their study on the formation of subcultures on TikTok to collect 165,000 videos. These studies illustrate the capacity of the API to generate larger samples than UI-based methods have been able to produce. Moreover, because the TikTok Research API operates independently of user accounts, the resulting datasets are not shaped by algorithmic curation based on user-level behaviors.

However, scholars have identified several limitations of the research API. For instance, \citet{corso_what_2024}found regional and temporal biases in the Research API dataset based on 577,517 videos retrieved across all regions over six years. They showed that the majority of data consists of videos posted on the first day of the month, with most coming from India and Southeast Asia \cite{corso_what_2024}. Similarly, Luceri et al. comment on the structural limitations of the Research API, including ``rate  limits,  incomplete  metadata  returns,  and  inconsistent quota  fulfillment'' \cite{luceri2026coordinated}. Because of these limitations, Luceri et al. accounted for data loss by using third-party data collection tools and crawling infrastructure. These limitations indicate that the Research API does not always return data in line with the platform's documentation. Another study compared videos returned by the API to web data and found that some publicly available videos are missing from API results \cite{pearson_beyond_2025}. These discrepancies may introduce critical limitations for studies conducting content analysis on TikTok metadata, especially for those whose goal is to explain how a certain event is discussed on the platform and which topics are most often seen by end-users.

\section{Data Collection}

\begin{figure*}[t]
    \centering
    \includegraphics[width=\textwidth]{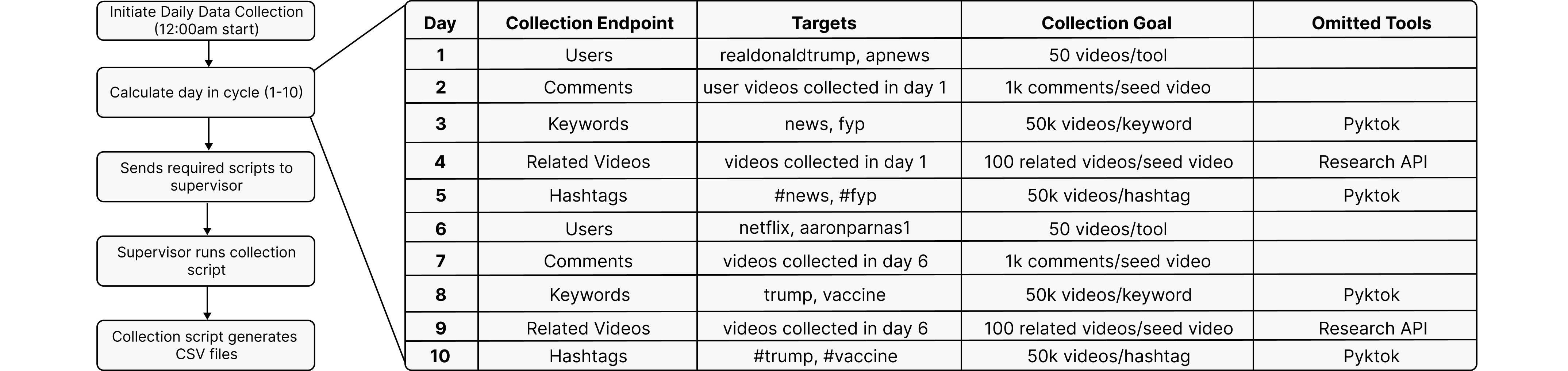}
    \caption{Automated data collection workflow showing the daily execution schedule and 10-day rotating endpoint cycle.}
    \label{fig:workflow}
\end{figure*}

To empirically examine the issues discussed above, we conducted a comparison study of three different TikTok data collection tools---Apify, Pyktok, and TikTok's official Research API---focusing on (1) which tools collect the most data by endpoint and (2) data overlap across tools for identical queries. To facilitate data access over longer time intervals, we queried the TikTok Research API via its official Python wrapper.\cite{tiktok_research_api_wrapper} While the API restricts each video query to a 30-day window, the wrapper automated repeated queries across multiple 30-day periods. All tools and APIs used in this study are publicly available.

We attempted to standardize the data collection procedures for each tool to ensure that their datasets could be meaningfully compared. We implemented a fully automated longitudinal data collection pipeline that enforced consistent scheduling and query execution across all collection methods (\textit{see Figure~\ref{fig:workflow}}). Data were collected over a 30-day period (August 1-30, 2025), divided into three repeated 10-day collection cycles. A daily cron job triggered at 00:00 UTC executed the collection scripts corresponding to the endpoint scheduled for that day. Each endpoint was executed twice per 10-day cycle at five-day intervals.

We tracked the same users, hashtags, keywords, and videos for 30 days.\footnote{During data collection, Pyktok’s hashtag queries were temporarily unavailable due to internal HTML changes requiring ongoing updates. Still, both Pyktok and Apify use browser-based scraping that better reflects user experience than API data, with Pyktok offering free access and Apify requiring a subscription.} Query terms were selected to cover domains and actors commonly studied in web and computational social science, including politics, public health, entertainment, and viral content. User queries included an elected official (\textit{@realdonaldtrump}), a news organization (\textit{@apnews}), an influencer (\textit{@aaronparnas1}), and a brand account (\textit{@netflix}). Keyword and hashtag queries included issue terms (\textit{trump, vaccine}) and widely used viral terms (\textit{news, fyp}).

For the TikTok Research API, we implemented two collection modes based on the hyperparameter \texttt{is\_random}. According to the official API documentation, when \texttt{is\_random} is \texttt{False} (the default), results are returned in decreasing order of video IDs, whereas when \texttt{is\_random} is \texttt{True}, the API returns matching videos in random order. \cite{tiktok_developers_research_api} The video query endpoint also requires explicit \texttt{start\_date} and \texttt{end\_date} values, and each request is limited to a date range of no more than 30 days. \cite{tiktok_developers_research_api} Because we aimed to cover the period since TikTok's U.S. launch, a single query over the full study period was not feasible. In addition, requests covering lengthy time intervals returned an \texttt{internal\_error} (HTTP 500; see \cite{tiktok_api_v2_error_handling}). We therefore adopted two different retrieval strategies:

\begin{enumerate}
    \item With \textit{\texttt{is\_random=False}}, we applied a 30-day sliding window, paging backward through time until reaching the target sample size (\textit{N=50,000 per query}) or the earliest date available (\textit{June 1, 2021}). We refer to this condition as API(D), where D stands for \textit{decreasing}.
    
    \item With \textit{\texttt{is\_random=True}}, we stratified the period into monthly intervals from each collection date back to the same June 1, 2021 temporal boundary, under the same 50,000-video cap. Each query used \textit{\texttt{max\_count=100}}, \textit{\texttt{max\_total=1000}}, and \textit{\texttt{fetch\_all\_pages=True}}, allowing up to 1,000 videos per stratum. We refer to this condition as API(R), where R stands for \textit{random}.
\end{enumerate}

These two modes served different retrieval purposes. API(D) was the primary collection mode and was designed to maximize data volume by paging through ordered results within successive 30-day windows. API(R), by contrast, was included as a diagnostic comparison condition and was designed to distribute random-mode queries across the historical period under the API's bounded date-query requirements. Monthly stratification was therefore used to ensure temporal coverage when examining the practical consequences of \texttt{is\_random=True}. Because API(R) was introduced only in the second data collection cycle, direct comparisons between API(D) and API(R) were restricted to matched observations, defined as the same search type, keyword, and collection date with nonzero records in both modes.

Finally, the daily collection goal was set to stress test the TikTok Research API at its maximum capacity of 1{,}000 daily requests. Each request retrieves up to 100 videos or comments. For \textit{Hashtag}, \textit{Keyword}, and \textit{Comment} endpoints, we allocated 500 requests per target, querying two targets daily. For \textit{User}, we collected the 50 most recent videos per user. For \textit{Related Video}, we collected 100 related videos per seed video. We requested equivalent amounts of data for the other tools. Under this design, the collection yields a theoretical target volume of approximately 600 user videos, 60{,}000 related videos, and 600{,}000 results for the hashtag, keyword, and comment endpoints, respectively.

{\footnotesize
\[
\begin{aligned}
600{,}000 &= (2\,\text{targets/d}) (6\,\text{d}) (500\,\text{reqs/target}) (100\,\text{results/req}) \\
60{,}000 &= (2\,\text{targets/d}) (50\,\text{seed vids/d}) (6\,\text{d}) (100\,\text{related vids/seed}) \\
600 &= (2\,\text{users/d}) (6\,\text{d}) (50\,\text{vids/user})
\end{aligned}
\]
}

\section{Data Analysis}
We report both tool-specific collection results and cross-tool comparisons. Our analysis focuses on divergences across data collection strategies under matched query conditions. This focus reflects the real-world constraints of TikTok data collection. Researchers do not have access to platform-level population data, the Research API imposes strict quota limits, and browser-based tools do not provide auditable sampling procedures. Under these conditions, the central methodological question is not which tool most closely reproduces the full platform population, since that benchmark is unavailable to independent researchers. Rather, the relevant question is whether existing collection strategies systematically retrieve different subsets of content. 

Tool-specific analyses focus on completeness and internal consistency. To assess coverage, we employ a success rate metric defined as ($\frac{N_{\text{collected}}}{N_{\text{goal}}}$). A value of 1 indicates that a tool reached the daily collection target.

Cross-tool comparisons examine overlap in retrieved content, as well as differences in volume, user characteristics, and engagement metrics. To quantify overlap, we use Jaccard similarity coefficients ($J$) for each content type under the same query and quota conditions. This measure captures the extent to which different tools return the same content when given identical inputs. Engagement metrics are compared using Wilcoxon rank-sum tests because these measures are right-skewed in our data and the underlying sampling processes are quota-constrained and not fully observable. 


\section{Collection Success Rates by Tools}

\begin{table}[t]
\centering
\footnotesize
\setlength{\tabcolsep}{4pt}
\begin{tabular}{lrrrrr}
\hline
Tool & User & Hashtag & Keyword & Comments & Related \\
\hline
\textit{Goal} & \textit{600} & \textit{600K} & \textit{600K} & \textit{600K} & \textit{60K} \\
\hline
API(D)  & 873 & 442K & 419K & 233K & -- \\
        & (146\%) & (74\%) & (70\%) & (39\%) &  \\
Pyktok  & 921 & -- & -- & 349K & 9.1K \\
        & (154\%) &  &  & (58\%) & (15\%) \\
Apify   & 700 & 10K & 3K & 360K & 19K \\
        & (117\%) & (2\%) & (0\%) & (60\%) & (32\%) \\
\hline
\end{tabular}
\caption{Data collection success rates across tools and endpoints, measured as the percentage of predefined collection goals achieved.}
\label{tab:success}
\end{table}

\begin{figure}
  \centering
  \includegraphics[width=\columnwidth]{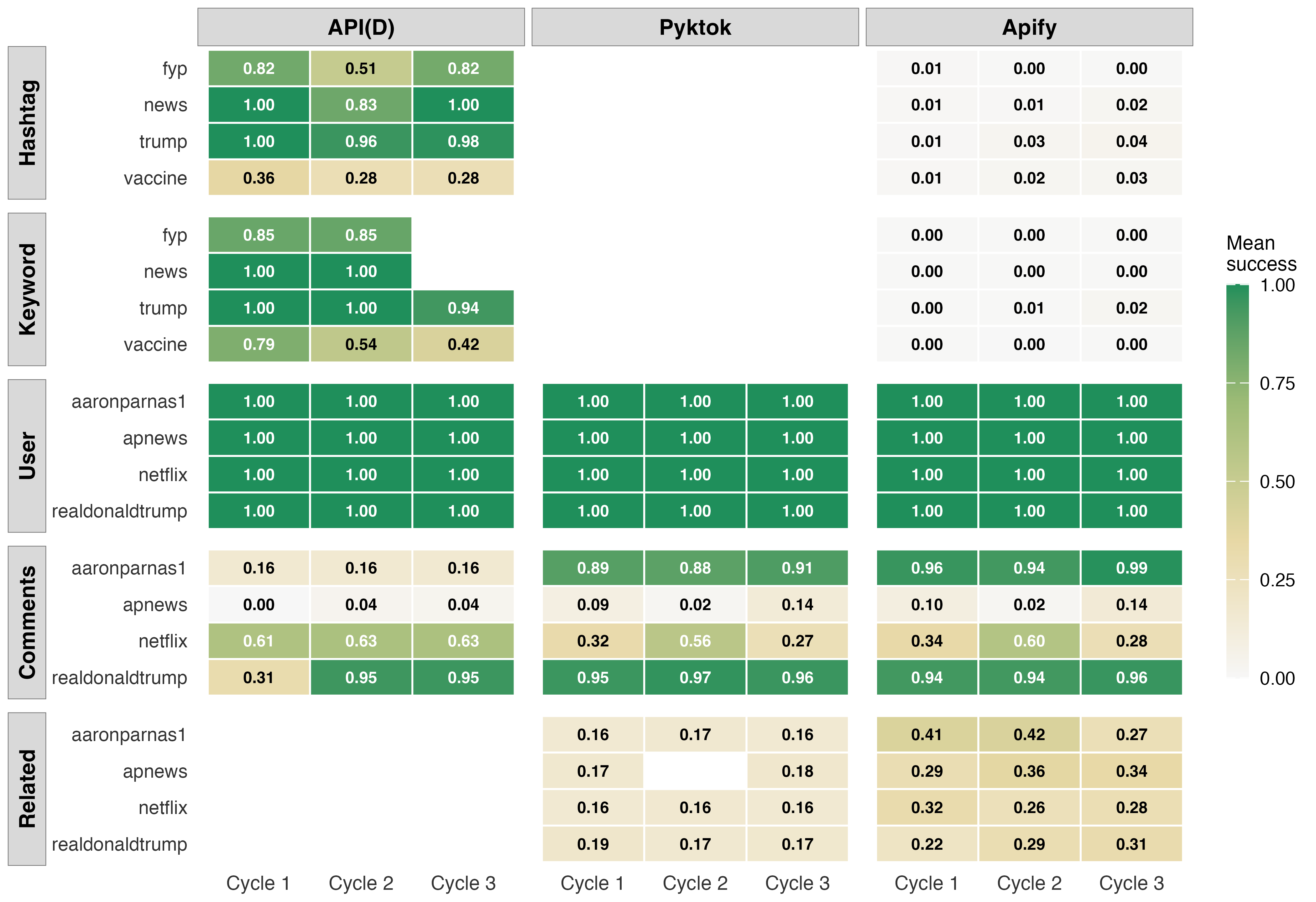}
  \caption{Heatmap of data collection success rates across tools, endpoints, targets, and collection cycles. Success rate is defined as the ratio of collected items to predefined collection goals, \(N_{\text{collected}} / N_{\text{goal}}\), capped at 1.}

  \label{fig:success}
\end{figure}

The collected data reveal substantial differences between tools (Table ~\ref{tab:success}). We begin with \textit{User} and \textit{Comments}, the two endpoints available across all three tools. The success rates for the \textit{User} endpoint were consistently equal to or greater than 1, indicating that more data were returned than requested. For the \textit{Comments} endpoint, Pyktok achieved the highest success rates, followed by Apify and API(D).

In contrast, substantial differences emerged in the hashtag and keyword endpoints. For hashtags and keywords, API(D) exhibited consistently higher goal ratios, whereas API(R) returned substantially lower ratios across queries (Table~\ref{tab:api_dr_matched}). Moreover, the two API tools differ in how individual queries contribute to the total. API(D) exhibits notable variation across keywords and hashtags, whereas API(R) consistently returned around 5,000 ($\sim 0.1$) videos per day. With API(D), the queries \textit{trump} and \textit{news} generally returned sample sizes close to their targets, with mean goal ratios near $\sim$0.98 and $\sim$0.97, respectively. The \textit{fyp} query had lower goal ratios, averaging $\sim$0.77, while \textit{vaccine} performed lowest overall, especially for hashtags. 

These results show that the volumes of data returned by the Research API varies depending on the value of the \textit{\texttt{is\_random}} parameter. In both settings, sample sizes frequently fall short of the targets described in the API documentation. Analysis of our collection logs indicates that, under \textit{\texttt{is\_random=False}}, most data loss—and the resulting variation in API(D) success rates—occurs when requests exceed the API’s rate limits during pagination. While this information is useful, managing this behavior remains challenging. The API does not support partial retrieval or early termination once a request is initiated, preventing researchers from stopping collection when a target quota is reached while retaining data collected up to that point. A plausible mitigation may be to reduce the queried time window. However, this approach is also limited, as the API provides no visibility into the total volume of available data within a given period. As a result, collection outcomes remain difficult to predict and are prone to data loss. In contrast, when \textit{\texttt{is\_random=True}}, the API consistently returns a fixed-size sample of approximately 10\% of the documented availability target. This behavior is not described in the publicly available API documentation.

Meanwhile, Apify showed substantially lower return rates for keywords and hashtags ($\sim 0.01$), with the query \textit{trump} achieving the highest average success rate (0.036), consistent with the Research API. Still, the remaining queries present a distinct hierarchy. \textit{vaccine} (0.021) outperformed both \textit{news} (0.013) and \textit{\#fyp} (0.011), contrasting with the Research API. However, our analysis suggests that this shortfall is unlikely to reflect bias in Apify itself; rather, it likely results from TikTok’s content curation, since Apify only scrapes front-end data.

\section{Data Coverage and Overlap by Tools}

We begin the comparative analysis by examining the \textit{User} and \textit{Comments} endpoints, which were available across all three tools. We then conduct paired comparisons between tools for endpoints that were available through only some of the tools.

\subsection{All three tools: User \& Comments Endpoints}

\begin{table}[t]
\centering
\footnotesize
\setlength{\tabcolsep}{4pt} 
\begin{tabular}{lcccc}
\hline
Endpoint & API(D) vs & API(D) vs & API(D) vs & Apify vs \\
         & API(R)    & Apify     & Pyktok    & Pyktok   \\
\hline
Hashtag  & 0.00 & 0.00 & ---  & ---  \\
Keyword  & 0.00 & 0.00 & ---  & ---  \\
User     & ---  & 0.03 & 0.06 & 0.71 \\
Comments & ---  & 0.06 & 0.00 & 0.00 \\
Related  & ---  & ---  & ---  & 0.00 \\
\hline
\end{tabular}
\caption{Pairwise Jaccard similarity coefficients between tools for each collection endpoint, measuring overlap in the sets of collected items. Dashes indicate endpoint–tool pairs that were not supported or not comparable.}
\label{tab:overlap}
\end{table}

\paragraph{User Endpoint}

Although the \textit{User} endpoint exhibits comparatively higher overall similarity, the Research API diverges from Pyktok and Apify (Jaccard similarity $J = 0.06$ and $J = 0.03$, respectively), whereas Pyktok and Apify are substantially more similar to each other ($J = 0.71$). This makes sense given that the two scraping-based platforms collect data via TikTok's web front-end, while the Research API does not. API(D)'s only shared videos with Pyktok and Apify were @realdonaldtrump's videos. For @realdonaldtrump, API(D) and Pyktok show significantly more overlap with each other than either does with Apify (Table ~\ref{tab:trump_overlap}). This difference corresponds to observed differences in temporal coverage, with Pyktok and the Research API collecting videos created between 2024-06-02 and 2024-11-05, while Apify’s data extend back only to 2024-08-05.

\begin{table}[t]
\centering
\footnotesize
\setlength{\tabcolsep}{3pt}
\resizebox{\columnwidth}{!}{%
\begin{tabular}{lrrrrrrr}
\hline
 & \multicolumn{4}{c}{Counts} & \multicolumn{3}{c}{Overlap} \\
\cline{2-5}\cline{6-8}
Cycle & \shortstack{Total\\Unique} & Apify & Pyktok & API(D) &
\shortstack{Apify-\\Pyktok} &
\shortstack{Apify-\\API(D)} &
\shortstack{Pyktok-\\API(D)} \\
\hline
Cycle 1 & 85 & 27 & 4 & 0 & 1 & 0 & 31 \\
Cycle 2 & 85 & 27 & 5 & 0 & 1 & 0 & 30 \\
Cycle 3 & 85 & 27 & 5 & 0 & 2 & 0 & 30 \\
\hline
\end{tabular}%
}
\caption{Per-cycle user counts and cross-tool overlap for \textit{@realdonaldtrump} across data collection tools.}
\label{tab:trump_overlap}
\end{table}

For @aaronparnas1, @apnews, and @netflix, Apify collected 50 videos for each user consistently across cycles, all of which were also captured by Pyktok. Still, Pyktok collected 20-29 additional videos beyond Apify's 50 videos. The shared user videos' time frame falls between 2025-07-29 and 2025-08-03, with Pyktok's additional videos extending the temporal range slightly earlier, capturing 20-29 videos from late July 2025 that Apify did not retrieve.

\paragraph{Comments Endpoint}

\begin{figure}[t]
    \centering
    \includegraphics[width=\columnwidth]{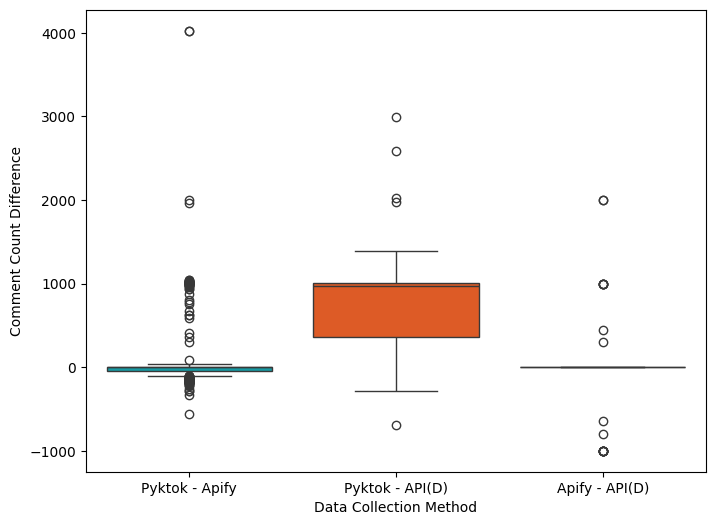}
    \caption{Distribution of pairwise differences in comment counts between data collection tools for hashtag- and keyword-based queries. The circles represent outliers, or extreme differences, in the number of comments collected by different tools for the same video.
    }
    \label{fig:comment}
\end{figure}

For the \textit{Comments} endpoint, Pyktok consistently returned more comments than API(D) for the same videos (\textit{M} = 801; \textit{SD} = 703). Approximately 40\% of comments overlapped between Pyktok and API(D) for shared videos (Figure~\ref{fig:comment}). By contrast, comparisons involving Apify yielded similar quantities of comments for shared videos, yet no overlapping comments were observed in either the Pyktok–Apify or Apify–API(D) comparisons.

These comparisons were conducted under the official API constraint of a maximum of 1,000 retrievable comments per video. Within this fixed limit, the absence of overlap in comparisons involving Apify indicates substantial differences in comment selection between tools. In particular, Apify did not show overlap with either Pyktok or API(D), indicating a distinct comment retrieval process that raises concerns for reproducibility.

\subsection{API(D) vs API(R)}

To reiterate, with API(D), the \textit{is\_random} parameter is set to False, meaning that it returns results in decreasing order of video IDs. API(R) means that the parameter is set to True, thus generating random samples (according to TikTok).

\subsection{Observed Retrieval Behavior of the \textit{\texttt{is\_random}} Parameter}

We first report an observed retrieval behavior of the TikTok Research API wrapper that emerged during collection and was subsequently reproduced in a diagnostic test. API(R) returned substantially fewer videos than API(D) across matched hashtag and keyword queries (Table~\ref{tab:api_dr_matched}). API(D) generally approached the configured retrieval target for high-volume queries, with unique video counts ranging from 13,875 to 50,073 across the samples analyzed here. Collections for hashtag and keyword searches using \texttt{news} and \texttt{trump} often returned close to 50,000 unique videos. API(D) yielded fewer results for lower-volume or more constrained queries, particularly \texttt{vaccine}, where unique video counts ranged from 13,875 to 26,971 depending on search type and collection date.

API(R) sample sizes were much smaller. Across the API(R) samples analyzed in this study, unique video counts ranged from 354 to 7,028, with most collections containing roughly 4,000 to 5,000 records, or about 10\% of the target amount. This notable discrepancy warranted further examination of the wrapper's pagination behavior under \texttt{is\_random=True} and \texttt{fetch\_all\_pages=True}.

The diagnostic test reproduced this pattern. Under the same hashtag, date range, and page size, \texttt{is\_random=True} and \texttt{is\_random=False} produced different pagination behavior. For a \#fyp query covering February 1 through February 28, 2026, the \texttt{is\_random=True} request returned 88 videos and \texttt{has\_more=False}, so the wrapper stopped after the first page. In contrast, the otherwise identical \texttt{is\_random=False} request returned 84 videos with \texttt{has\_more=True} on the first page and continued paginating to more than 1,000 videos. Across the broader diagnostic test, \texttt{is\_random=True} responses did not return the continuation fields needed for pagination, with \texttt{has\_more=False}, \texttt{cursor=None}, and \texttt{search\_id=None}. As a result, API(R) collected only the first non-paginated response for each query, yielding roughly 180 to 200 unique videos in the diagnostic test. By contrast, \texttt{is\_random=False} often returned continuation fields that allowed the wrapper to paginate through additional result pages. Table~\ref{tab:api_random_pagination_replication} in the Appendix reports the replication test across hashtag and keyword targets.

The collected API(R) samples are consistent with this pagination behavior. Monthly stratification produced broader temporal coverage, but many monthly requests returned fewer than the 1,000-video ceiling specified in our collection parameters. Accordingly, API(R) sample sizes appear to reflect both the number of monthly strata queried and the number of records returned within each stratum.

\begin{table}[t]
\centering
\footnotesize
\setlength{\tabcolsep}{3pt}
\begin{tabular}{llcrrrr}
\toprule
Type & Query & Date &
\multicolumn{2}{c}{API(D)} &
\multicolumn{2}{c}{API(R)} \\
\cmidrule(lr){4-5} \cmidrule(lr){6-7}
& & & Count & Rate & Count & Rate \\
\midrule
Hashtag & fyp & 25-08-15 & 25,632 & 0.51 & 4,080 & 0.08 \\
Hashtag & fyp & 25-08-25 & 40,874 & 0.82 & 4,876 & 0.10 \\
Hashtag & news & 25-08-15 & 41,716 & 0.83 & 4,936 & 0.10 \\
Hashtag & news & 25-08-25 & 50,064 & 1.00 & 5,174 & 0.10 \\
Hashtag & trump & 25-08-20 & 47,852 & 0.96 & 4,195 & 0.08 \\
Hashtag & trump & 25-08-30 & 49,040 & 0.98 & 4,141 & 0.08 \\
Hashtag & vaccine & 25-08-20 & 14,079 & 0.28 & 4,987 & 0.10 \\
Hashtag & vaccine & 25-08-30 & 13,875 & 0.28 & 4,963 & 0.10 \\
Keyword & news & 25-08-13 & 50,073 & 1.00 & 3,733 & 0.07 \\
Keyword & trump & 25-08-18 & 50,017 & 1.00 & 6,043 & 0.12 \\
Keyword & trump & 25-08-28 & 47,159 & 0.94 & 4,053 & 0.08 \\
Keyword & vaccine & 25-08-18 & 26,971 & 0.54 & 7,028 & 0.14 \\
Keyword & vaccine & 25-08-28 & 20,906 & 0.42 & 5,127 & 0.10 \\
\midrule
Total &  & 13 & 478,258 &  & 63,336 &  \\
\bottomrule
\end{tabular}
\caption{Matched API(D) and API(R) collection observations used in the diagnostic comparison. Only matched endpoint, query, and collection-date pairs are included. Counts are unique video IDs, and success is calculated relative to the 50{,}000-video target for each query.}
\label{tab:api_dr_matched}
\end{table}

\subsubsection{Hashtag \& Keyword Endpoints}

\paragraph{Temporal coverage}
{The temporal range of videos collected by API(D) varied substantially by query. API(D) often returned videos from narrow recent windows, especially for high-volume queries. For example, \textit{fyp} spanned 6 minutes in the keyword endpoint, while \textit{\#fyp} ranged from 8 minutes to 1.9 months in the hashtag endpoint. Lower-volume queries reached farther back in time, with \textit{\#vaccine} spanning approximately 3 to 5 months and \textit{vaccine} spanning about 28 days to 4 months (Table~\ref{tab:temporal_span}). API(R) data spans averaged 50 months across queries, reflecting successful monthly stratification and broader temporal coverage than API(D).

\paragraph{Engagement metrics}
Both API(D) and API(R) showed extremely right-skewed distributions with relatively low engagement metrics across all queries. The differences between their engagement metrics showed very small effect sizes (Figure~\ref{fig:hashtag_keyword_engagement_dist}).

\begin{figure}[]
    \centering
    \includegraphics[width=1\columnwidth]{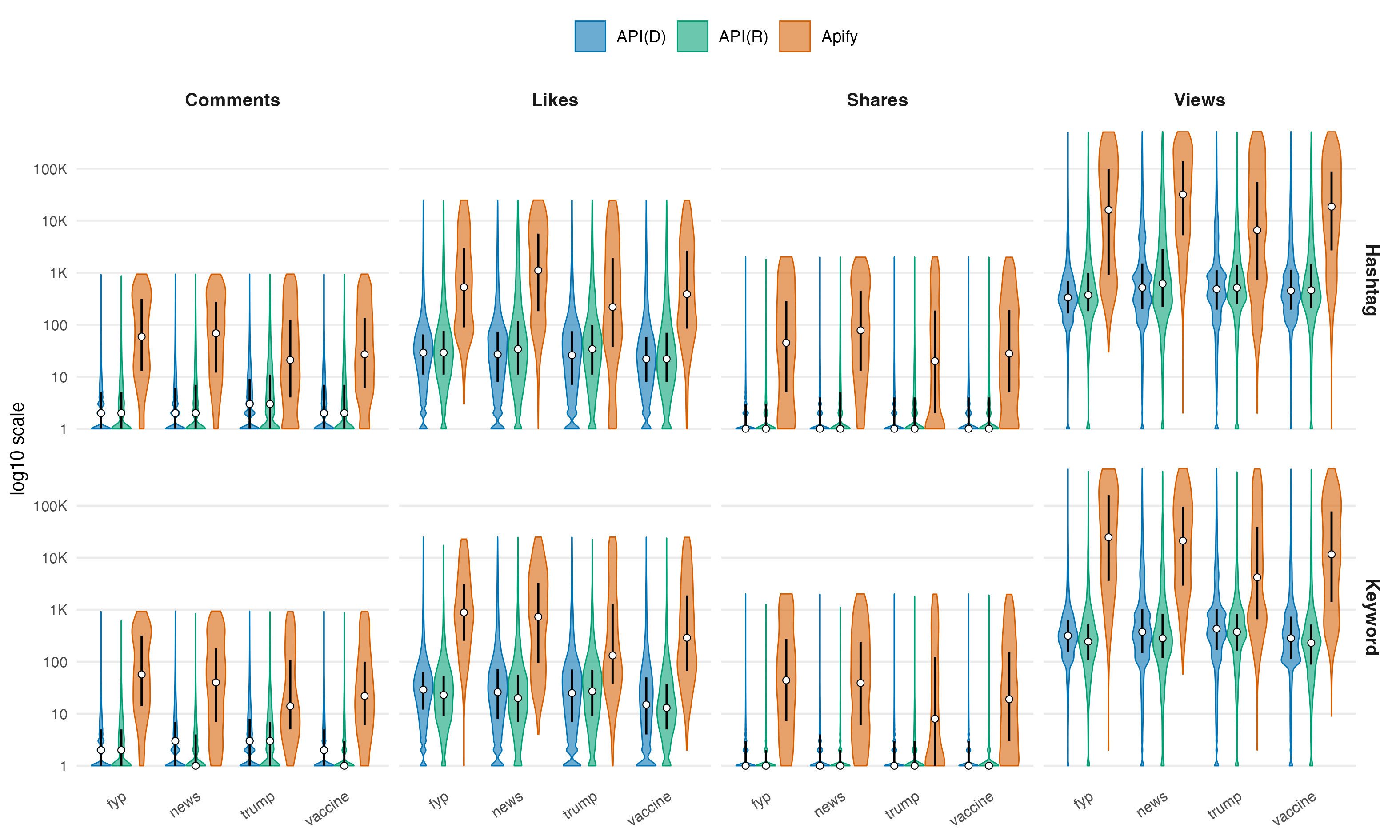}
    \caption{Engagement distributions across collection tools, endpoints, and query targets. Columns show comments, likes, shares, and views; rows distinguish hashtag and keyword endpoints. The y axis is log10 transformed after adding one to each count. White points mark medians, and black lines show interquartile ranges. API(D) and API(R) show similar engagement profiles, while Apify returns videos with higher engagement.}

    \label{fig:hashtag_keyword_engagement_dist}
\end{figure}

Without access to TikTok's complete video population or its underlying architecture, researchers cannot independently verify whether samples drawn from API(R) are truly ``random" as typically defined. We therefore do not present the API(D) and API(R) comparison as a synchronized experiment designed to identify the direct impact of \texttt{is\_random}. Instead, we treat API(R) as a diagnostic comparison condition that describes the practical behavior of this wrapper setting under our collection design.

Still, two descriptive patterns are evident. First, there was no overlap between the videos returned by the two modes ($J=0.00$) for both \textit{Keyword} and \textit{Hashtag} endpoints. Second, despite returning distinct videos, API(D) and API(R) did not differ significantly in engagement or user metrics. As shown in Figure~\ref{fig:hashtag_keyword_engagement_dist}, the distributions of comment, like, share, and view counts are closely aligned across hashtags and keywords, indicating that both official API modes returned videos with similar engagement profiles.

Engagement on social media platforms is typically highly skewed, with a small number of videos receiving very high engagement and most videos receiving relatively little engagement. Thus, if the data population accessible by the API is dominated by lower engagement videos (as the full population likely is), both retrieval modes may reasonably return similar engagement distributions. What we observe is more specific. API(D) and API(R) returned non-overlapping sets of video IDs ($J=0.00$) for both \textit{Keyword} and \textit{Hashtag} endpoints, yet their engagement and account profile distributions were statistically similar. Under our collection design, enabling \texttt{is\_random} changed the specific videos retrieved but did not materially change the aggregate engagement profile of the Research API output.

However, when compared with Apify, we observe a distinct divergence. In the next section, we show that the API's stable low engagement profile differs sharply from the videos collected through Apify, suggesting that the conclusions researchers draw about TikTok content can depend substantially on the data collection tool employed.

\subsection{API(D) vs Apify}

Despite controlling for potential confounding factors, including the date and time of collection, collection window length, and API rate limits, we observe no overlap between the Apify and Research API datasets for hashtag and keyword endpoints ($J = 0.00$).\footnote{Our analysis confirms that both datasets contain videos from overlapping time periods, ruling out differences in content creation dates as an explanation for the zero Jaccard coefficients.}

\subsubsection{Hashtag \& Keyword Endpoints}

\paragraph{Temporal coverage}

\begin{table}[t]
\centering
\footnotesize
\setlength{\tabcolsep}{3pt}
\renewcommand{\arraystretch}{0.82}
\begin{tabular}{llcc}
\toprule
Endpoint & Query & API(D) & Apify \\
\cmidrule(lr){3-3}\cmidrule(lr){4-4}
 & & Observed span & Observed span \\
\midrule
Hashtag & \#fyp     & 8 min--1.9 mo    & 67.9--80.3 mo \\
Hashtag & \#news    & 1.3 d--1.9 mo    & 57.3--65.2 mo \\
Hashtag & \#trump   & 2.7 d--3.0 mo    & 66.1--75.9 mo \\
Hashtag & \#vaccine & 3.8--4.7 mo      & 63.3--70.7 mo \\
\midrule
Keyword & fyp       & 6 min            & 59.5--80.2 mo \\
Keyword & news      & 11.7--13.5 h     & 28.7--54.8 mo \\
Keyword & trump     & 1.1 d--2.9 mo    & 60.0--64.6 mo \\
Keyword & vaccine   & 28.5 d--4.1 mo   & 58.3--64.8 mo \\
\bottomrule
\end{tabular}
\caption{Temporal coverage of returned videos across hashtag and keyword queries. For each query execution, span is calculated from video creation timestamps as the difference between the earliest and latest returned video. Table cells report the shortest and longest observed spans across collection cycles.}
\label{tab:temporal_span}
\end{table}

Overall, Apify data covered extended periods, typically spanning several years within each collection cycle. Specifically, the keyword \textit{news} spans $\sim$4 years, while the hashtag \textit{\#fyp} spans $\sim$6 years and reaches up to $\sim$7 years. This differs from API(D), which showed query-dependent temporal coverage and often returned videos from much narrower recent windows.

\paragraph{Engagement metrics}
Figure~\ref{fig:hashtag_keyword_effect_size} reports rank-biserial correlations from Wilcoxon rank-sum comparisons of engagement metrics across tools. The largest divergences appear in comparisons between Apify and both API conditions. Across comments, likes, shares, and views, API(D)-Apify and API(R)-Apify comparisons show consistently large effects, indicating that Apify returns videos with substantially higher engagement than either Research API condition. In contrast, API(D)-API(R) comparisons produce much smaller effects, suggesting that the two Research API modes are more similar to each other than either is to Apify. Metric-specific distributional plots are provided in Appendix Figures~\ref{fig:appendix_comment_ridgeline}--\ref{fig:appendix_view_ridgeline}, which show the same pattern at the distributional level.

Without ground truth data from TikTok, the observed pattern should be interpreted as divergence across retrieval strategies rather than as direct evidence that any one tool is biased relative to the full TikTok population. The differences could reflect Apify retrieving more highly engaged content, the Research API underrepresenting highly engaged content, or both samples deviating from the underlying population in different ways. The results nevertheless show that collection strategy strongly shapes the engagement distributions available for analysis.

\begin{figure}[t]
    \centering
    \includegraphics[width=\columnwidth]{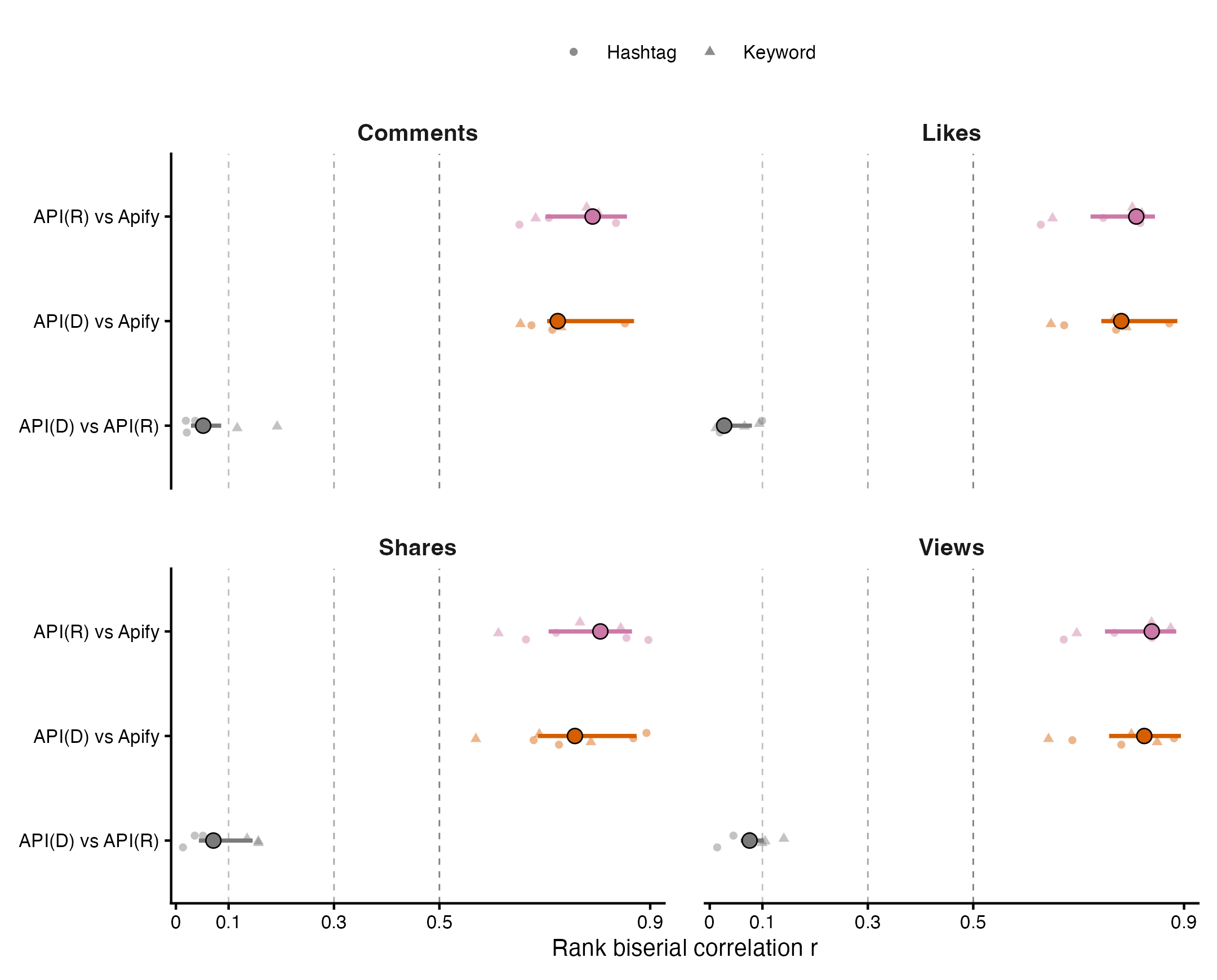}
    \caption{Rank-biserial effect sizes from pairwise comparisons of engagement distributions across collection tools. Small points represent endpoint-target comparisons, with circles for hashtag queries and triangles for keyword queries. Larger outlined points show median effects, with horizontal lines indicating the middle 50\% of effect sizes.}
    \label{fig:hashtag_keyword_effect_size}
\end{figure}

\paragraph{Data Source and Collection Results}
One notable empirical difference that may help account for these discrepancies lies in the data sources through which content is retrieved. Through logging, we discover that Apify operates through incognito browser-based scraping of TikTok's web interface, collecting data from either hashtag pages (e.g., \url{www.tiktok.com/tag/news}) or search result pages (e.g., \url{www.tiktok.com/search?q=news}). In other words, Apify scrapes what TikTok decides to show to end users. The low retrieval rates, therefore, likely stem from the limited data displayed on these pages compared to the underlying data pool to which the API has access. 

Taken together, our experiments with different data acquisition tools show massive divergences among collection strategies rather than direct evidence of bias in any single source. The API based collections returned many lower engagement videos and higher retrieval volumes, while Apify returned smaller collections with substantially higher engagement. These divergences matter because the conclusions researchers draw about TikTok content, accounts, and engagement can depend on the retrieval tool used to construct the dataset, rather than representing the platform as a whole.

This finding also relates to the methodological reporting issues noted in our literature review. These omissions are likely not intended to obscure data collection decisions. However, our results show that small choices in collection tools, query settings, and retrieval environments can produce nontrivial differences in the data researchers obtain. These choices have not always been fully explicit in current practice. Researchers often make data collection decisions within tools and interfaces provided by platform companies or third party services, each of which imposes its own defaults, quotas, pagination rules, query constraints, ranking logics, and failure modes. As a result, researchers face many degrees of freedom both in the choices they make and in the constraints they are forced to work within when using available data access tools.

Because TikTok data are cross-sectional and the platform changes continuously, including changes in user profiles, account handles, available videos, and interface behavior, exact replication of prior collections may often be difficult. This makes transparent reporting even more important. Researchers should document not only the data source, but also the collection tool, endpoint or interface, query type, search terms, date ranges, timing of collection, pagination settings, randomization settings, quotas, stopping rules, and any observed collection failures or retries. Such reporting would improve the interpretability and reproducibility of TikTok studies, while also clarifying what conclusions a given study can and cannot support.

\subsection{Hourly vs.\ Four-Hour Front-End Scraping Intervals}

\begin{figure}[t]
    \centering
    \includegraphics[width=0.9\columnwidth]{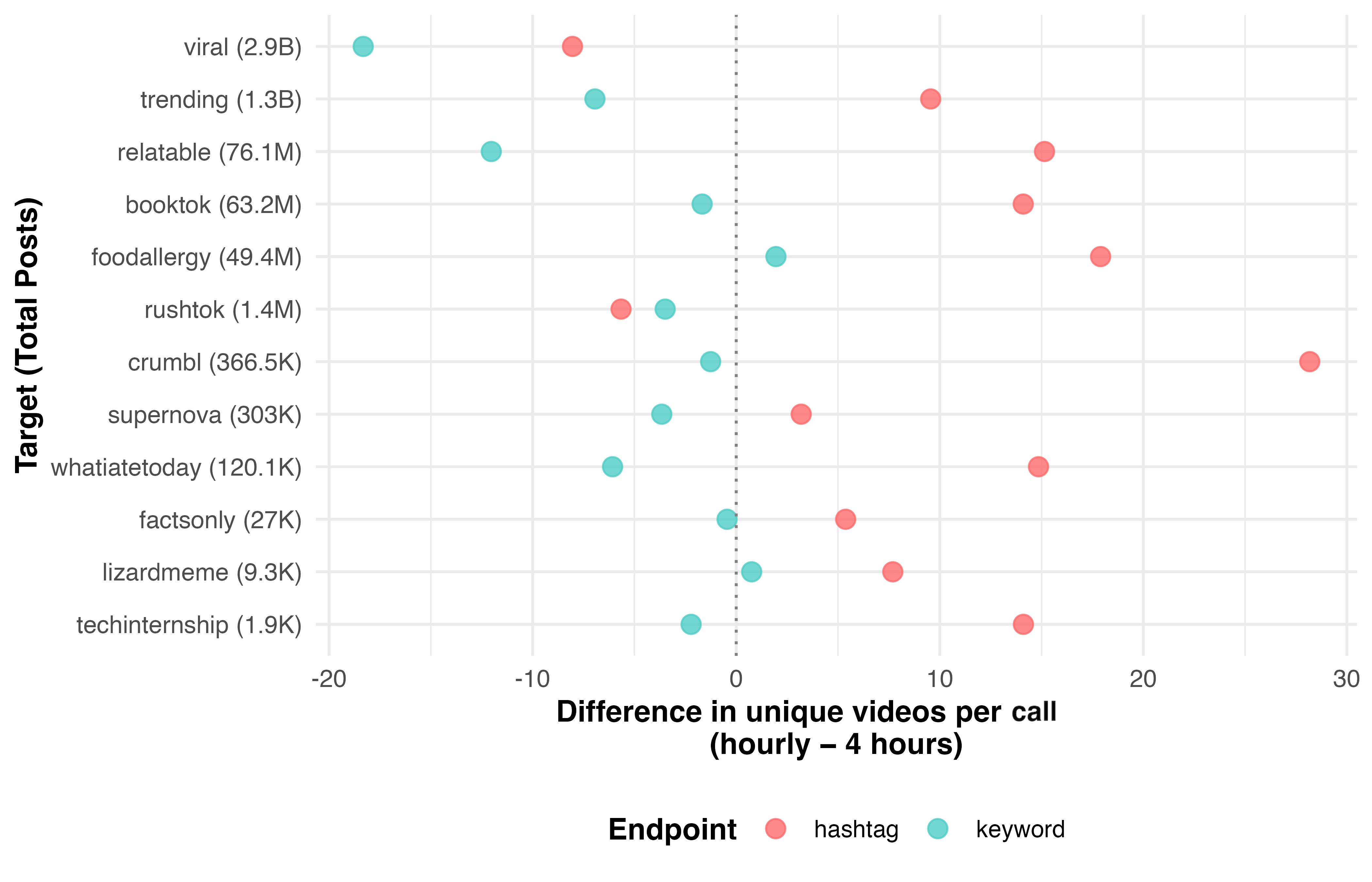}
    \caption{Each point represents the difference in unique videos scraped per session when using hourly intervals versus 4-hour intervals. Positive values indicate that hourly collection yields more unique videos per call. Targets are ordered by total post volume (shown in parentheses, retrieved from TikTok tag page metadata at tiktok.com/tag/\{target\} as of data collection)}
    \label{fig:optimal_scraping_interval}
\end{figure}

While web scraping yielded substantially fewer results than API based collection, it has a different methodological value because it more closely approximates the content visible through TikTok's web interface. This difference raises two related methodological issues for scraper based collection. The first concerns whether increasing collection frequency, by shortening the interval between consecutive scraping sessions, improves the number of newly retrieved unique videos per request. The second concerns whether retrieval yield varies with the popularity of the target term, as measured by the total number of posts associated with the hashtag or keyword.

We therefore conducted an additional scraping frequency test using targets that varied widely in platform level post volume, from low volume tags with fewer than 2,000 posts to highly popular tags with billions of posts. This allows us to examine whether the observed collection dynamics are specific to high volume targets or whether they also appear across terms with different levels of platform prevalence.

We compared hourly and 4 hour scraping intervals across these targets. Figure~\ref{fig:optimal_scraping_interval} reports two front-end scraping intervals---hourly collection (P1), which represents a tighter sampling strategy, and collection every four hours (P2), which represents a looser strategy. For each target \(i\), we compute the difference in per-request yield as
\[
\Delta N_i = N_i^{(1h)} - N_i^{(4h)},
\]
where \(N_i^{(1h)}\) and \(N_i^{(4h)}\) denote the number of unique videos retrieved per request under the hourly and four-hour intervals, respectively. Positive values of \(\Delta N_i\) indicate higher yields under the tighter hourly interval, whereas negative values indicate higher yields under the looser four-hour interval.

We evaluate this difference across 12 terms, each queried both as a hashtag and as a keyword, yielding 24 comparisons in total. As shown in Figure~\ref{fig:optimal_scraping_interval}, differences in per-request yield between the two scraping intervals are generally small and concentrated near zero for both hashtags and keywords, indicating limited sensitivity to scraping frequency. Targets are ordered along the vertical axis by overall content volume on TikTok, with more popular terms shown toward the top of the plot. Keyword queries more often exhibit modest negative values of \(\Delta N_i\), suggesting slightly higher yields under the looser four-hour interval, whereas hashtag queries more frequently show modest positive values, indicating marginally higher yields under tighter hourly scraping. However, these differences are inconsistent across targets and do not increase systematically with content popularity. 

\section{Discussion}

Concerns regarding the reproducibility of scientific research, first articulated decades ago by \citet{newell1973you} in the domain of Psychology and later reiterated by \citet{watts2017should} in the context of Computational Social Science, remain salient in contemporary research. These concerns are particularly relevant to web-based research, where data collection practices often depend on rapidly evolving platforms and tools. Our systematic TikTok scraper audit reveals similar transparency deficits, underscoring the need for academic researchers to systematically document their data collection methods and limitations. Greater transparency in these practices is essential for enabling researchers to assess the reliability, reproducibility, and comparability of empirical findings. To address these challenges, we offer two types of recommendations: transparency guidance and methodological guidance.

\subsection{Transparency guidance}

Overall, different TikTok data collection method yielded vastly, and at times completely, different data samples. Specifically, analyses based on data from TikTok's Official Research API produce different results compared to front-end data scraping methods like Pyktok and Apify, thus possibly leading to different empirical claims. For example, when studying user responses to TikTok videos, Pyktok consistently retrieves more comments than Apify or the Research API for the same videos. Moreover, different tools capture only 0--40\% of the same comments for any given video, meaning each tool may return a largely different dataset for the same queries. These variations in data collection could lead researchers to draw divergent conclusions about the nature and content of user engagement with TikTok videos.

Mirroring academic research on earlier social media platforms like YouTube \cite{chandio2024audit}, early empirical research on TikTok is hampered by a lack of knowledge about the platform and its available data collection methods. Furthermore, as many studies employ content analysis as a primary research method, it is essential that researchers provide a transparent description of their data collection procedures. Content analysis relies on statistical techniques to draw inferences, and its validity depends on the implementation of appropriate sampling procedures \cite{krippendorff_conceptual_2019}, with the ultimate goal of making replicable and valid inferences about observed content.

We strongly encourage researchers beginning or scoping their TikTok projects to be as transparent as possible regarding their data collection methods. First, previously published articles using empirical TikTok data should be reconsidered in light of the new information presented in this paper. Second, new studies should disclose data collection methods in detail and discuss the limitations of their samples. The studies by \cite{steinke_women_2024, lee_identification_2025, izquierdo-condoy_assessing_2025} exemplify minimal transparency practices, which include describing the sampling method and acknowledging that it may have limitations. We recommend that researchers go beyond these minimal practices by incorporating discussions of the strengths and weaknesses of the data collection tool(s) they used and the related implications for answering their research questions.

\subsection{Methodological guidance}
The massively divergent results of the three data collection tools reviewed in this study may leave some researchers wondering how to proceed empirically. The answers depend on various factors, including the nature of a given study's sampling frame and the types of content and metadata desired. Before we consider each tool's data capacity, we must first assess whether it can collect the kind of data we want in the first place. As of this writing, Pyktok is not capable of collecting data from hashtags or search keywords (although hashtag functionality is in the process of being restored), and the Research API cannot collect ``You May Like'' videos. Researchers who need to generate their samples based on these specific criteria can easily eliminate tools that are incompatible with them.

However, because each endpoint can be accessed by at least two tools, that still leaves some consequential decisions to be made. Starting with the user endpoint, we must first recall that individual users' videos are listed on their profile pages chronologically (except for pinned videos); therefore, the sole consideration is the total number of videos collected. By that metric, we recommend Pyktok, as it collected the most videos, followed by the Research API, and then Apify (see Table~\ref{tab:success}). Proceeding to hashtags and keywords, between the Research API and Apify, the former collected over 442{,}000 videos for hashtags and 419{,}000 for keywords, whereas the latter only collected around 10{,}000 for hashtags and 7{,}000 for keywords. However, as we discuss in the Results, the former is biased in favor of lower-engagement videos compared to the latter. Therefore, researchers should consider not only the total volume of data available, but also the kinds of videos and users represented in each set of results.

Those primarily interested in high-visibility content might choose Apify for hashtags and/or keywords, although its price and the immense disparity in data volumes compared to the Research API might sway others in the opposite direction. For comments, Pyktok came in substantially ahead of its competitors, so we recommend it for researchers who require large sample sizes. Finally, we note that Apify collected about double the number of ``You May Like'' videos as Pyktok. When cost is no object, Apify is the clear choice, but researchers on a budget could choose Pyktok.

The low to nonexistent overlap between different samples collected by different tools raises the possibility of combining them. But because the respective characteristics of the data produced by the scraping-based methods compared to the official API can differ so vastly, we do not recommend simply pooling multiple types of datasets and proceeding with analysis. One option for using data extracted from multiple sources to test a single hypothesis is to use data collection tool as a moderator variable. Doing so would allow researchers to ascertain the extent to which the relationships between the predictors and outcome variable change based on how the data was procured. In cases where model outcomes remain consistent across data collected by different tools (i.e. significant main effects but no moderator effects for the tool variable), researchers could claim stronger support for their results. This is especially important given that there is currently no way for independent researchers to draw truly random samples from TikTok, and it is thus impossible for them to generalize any sample-based findings to the platform as a whole.

\subsection{Ethical guidance}
Lastly, it is important to acknowledge platform compliance considerations, as the data collection methods examined in this study differ in their adherence to TikTok’s Terms of Service (ToS) and the legal and ethical risks they may entail.

TikTok's Research API is the platform's only ToS-compliant data collection tool. However, the API is only available to researchers based in the US and Europe. The Research API's Terms prohibit third parties from accessing TikTok Research Tools without TikTok's express written authorization. They further restrict how data obtained through the API can be used: namely, it may only be used for the specific project described in the original application, and cannot be repurposed for other research \cite{tiktok_tiktok_2026}. Researchers based outside the US and Europe, therefore, cannot access TOS-compliant data through the methods reviewed here, nor through datasets shared by colleagues from eligible regions. Participant-mediated approaches such as data donation fall outside the scope of this review but offer a complementary pathway that operates independently of API access restrictions \cite{wedel_platform_2026}.

ToS violations are not ipso facto unethical. U.S. courts have repeatedly found that web scraping public data is legally protected \cite{SandvigSummaryJudgment2020,MetaBrightData_OrderShortenTime2023}. Researchers outside the U.S. should verify local legal provisions. Also, ToS compliance should generally be treated separately from ethical research conduct, as ToS are generally constructed to protect corporations, not to facilitate quality and necessary scientific inquiry \cite{Freelon2018, Freelon2025}. Consequently, we encourage researchers to make prudent and informed decisions regarding these tradeoffs in light of their research objectives, institutional review requirements, and applicable legal contexts.

\subsection{Limitations and Future Work}

Our study design was financially constrained, as is true with any study. We chose to focus on four high-visibility users plus four keywords representing a diverse array of topics studied in computational social science: politics, news media, new media, and health. Although our analysis cannot speak to all possible users and keywords, especially low visibility cases, we are confident that results clearly reflect meaningful differences between data collection tools. We do not incorporate a ``ground truth'' dataset because independent researchers do not have access to anything like that. \cite{mousavi_auditing_2024}. Further, due to resource limitations, we only tested Pyktok, Apify, and the TikTok Research API; however, future work should compare results from our study to outputs from alternative data collection methods. Lastly, Pyktok's hashtag and keyword endpoints were unavailable during data collection, limiting the comparability across tools. Unfortunately, this is a limitation and consequence of free, open source, and front-end scraping techniques. Ultimately, we consider this paper a starting point for researchers to critically question outputs from quantitative data collection tools that might otherwise be considered objective, comprehensive, or representative reflections of TikTok content.

\section{Conclusion}

Our findings highlight that data collection tools (TikTok's Official Research API, Apify, and Pyktok) return vastly different results for the same queries. This research has broad implications for web and computational social science research. Specifically, we find that when using front-end scraping techniques (e.g., Apify and Pyktok), data is skewed by TikTok's algorithm. The Official Research API, in contrast, returns videos with significantly lower engagement regardless of whether \textit{is\_random} is set to True or False.

Further, due to the sheer quantity of data on TikTok, complete samples were only possible for the user endpoint (and even then, only for users with relatively small video totals). Because data collection strategies yield different datasets, we raise concerns about the validity, reproducibility, and therefore generalizability of TikTok research. To mitigate potential shortcomings of this scholarship, we offer methodological guidance for endpoint selection and opportunities to combine tools, as well as transparency guidelines for reporting TikTok data collection methods.

\bibliography{reference}

@online{apify_web_2025,
  title        = {Web Scraping Basics for {JavaScript} Developers},
  author       = {{Apify}},
  year         = {2025},
  url          = {https://docs.apify.com/academy/scraping-basics-javascript},
  urldate      = {2025-09-29},
  organization = {Apify},
  langid       = {english},
}

@online{fanpagekarma_professional_,
  title        = {FanpageKarma: Professional Social Media Analytics and Management},
  author       = {{FanpageKarma}},
  year         = {2025},
  url          = {https://www.fanpagekarma.com/},
  urldate      = {2025-09-16},
  organization = {FanpageKarma},
  langid       = {english},
}

@online{octoparse_web_,
  title        = {Octoparse: Web Scraping Tool and No-Code Web Crawlers},
  author       = {{Octoparse}},
  year         = {2025},
  url          = {https://www.octoparse.com/},
  urldate      = {2025-09-16},
  organization = {Octoparse},
  langid       = {english},
}

@online{apify_apify_2025,
  title        = {About Apify},
  author       = {{Apify}},
  year         = {2025},
  url          = {https://apify.com/about},
  urldate      = {2025-10-01},
  organization = {Apify},
  note         = {Founded in 2015, Apify builds a platform for web data extraction and automation solutions worldwide.},
}

@inproceedings{goanta2026great,
  title={The great data standoff: Researchers vs. platforms under the digital services act},
  author={Goanta, Catalina and Zannettou, Savvas and Kaushal, Rishabh and van de Kerkhof, Jacob and Bertaglia, Thales and Annabell, Taylor and Gui, Haoyang and Spanakis, Gerasimos and Iamnitchi, Adriana},
  booktitle={Proceedings of the International AAAI Conference on Web and Social Media},
  volume={20},
  number={1},
  pages={874--888},
  year={2026}
}

@online{tiktok_research_2025,
  title = {Research {{API}}},
  author = {TikTok},
  year = {2025},
  url = {https://developers.tiktok.com/products/research-api/},
  urldate = {2025-09-03},
  organization = {TikTok for developers}
}

@online{adobe_express_using_2024,
  title = {Using {{TikTok}} as a {{Search Engine}}},
  author = {Adobe Express},
  year = {2024},
  url = {https://www.adobe.com/express/learn/blog/using-tiktok-as-a-search-engine},
  urldate = {2025-09-02},
  langid = {american},
  organization = {Adobe.com}
}

@online{diep_tiktok_2025,
  title = {{{TikTok}} as a {{Search Engine}}},
  author = {Diep, Pham Phuong Uyen and Tran, Huu Dat},
  year = {2025},
  eprinttype = {SSRN},
  url = {https://ssrn.com/abstract=5214981},
  langid = {english},
  pubstate = {prepublished}
}

@inproceedings{boeker_empirical_2022,
  title = {An {{Empirical Investigation}} of {{Personalization Factors}} on {{TikTok}}},
  booktitle = {Proceedings of the {{ACM Web Conference}} 2022},
  author = {Boeker, Maximilian and Urman, Aleksandra},
  date = {2022-04-25},
  year = {2022},
  pages = {2298--2309},
  publisher = {ACM},
  location = {Virtual Event, Lyon France},
  doi = {10.1145/3485447.3512102},
  url = {https://dl.acm.org/doi/10.1145/3485447.3512102},
  urldate = {2025-08-31},
  eventtitle = {{{WWW}} '22: {{The ACM Web Conference}} 2022},
  isbn = {978-1-4503-9096-5},
  langid = {english}
}

@inproceedings{pinto_fighting_2024,
  title = {Fighting for {{Democracy}}: {{The Attempted Coup}} in {{Peru}} through the Lens of {{TikTok}}},
  shorttitle = {Fighting for Democracy},
  author = {Pinto, Gabriel and Burghardt, Keith and Lerman, Kristina and Ferrara, Emilio},
  date = {2024},
  year = {2024},
  booktitle = {Proceedings of the {{ICWSM Workshop}} on {{Data}} for the {{Wellbeing}} of {{Most Vulnerable}}},
  eventtitle = {ICWSM Workshop on Data for the Wellbeing of Most Vulnerable},
  langid = {english}
}

@article{mousavi_auditing_2024,
  title = {Auditing {{Algorithmic Explanations}} of {{Social Media Feeds}}: {{A Case Study}} of {{TikTok Video Explanations}}},
  shorttitle = {Auditing {{Algorithmic Explanations}} of {{Social Media Feeds}}},
  author = {Mousavi, Sepehr and Gummadi, Krishna P. and Zannettou, Savvas},
  date = {2024-05-28},
  year = {2024},
  journaltitle = {Proceedings of the International 26
                  AAAI Conference on Web and Social Media},
  shortjournal = {ICWSM},
  volume = {18},
  pages = {1110--1122},
  issn = {2334-0770, 2162-3449},
  doi = {10.1609/icwsm.v18i1.31376},
  url = {https://ojs.aaai.org/index.php/ICWSM/article/view/31376},
  urldate = {2025-08-31},
  langid = {english}
}

@incollection{krippendorff_conceptual_2019,
  title = {Conceptual {{Foundation}}},
  booktitle = {Content {{Analysis}}: {{An Introduction}} to {{Its Methodology}}},
  author = {Krippendorff, Klaus},
  year = {2019},
  pages = {24},
  publisher = {SAGE Publications, Inc.},
  doi = {10.4135/9781071878781},
  url = {https://methods-sagepub-com.proxy.library.upenn.edu/book/mono/content-analysis-4e/toc},
  urldate = {2025-08-30},
  isbn = {978-1-0718-7878-1},
  langid = {english}
}

@misc{favero_teens_2024,
	title = {Teens, {Social} {Media} and {Technology} 2024},
	url = {https://www.pewresearch.org/internet/2024/12/12/teens-social-media-and-technology-2024/},
	language = {en-US},
	urldate = {2025-07-16},
	journal = {Pew Research Center},
	author = {Favero, Michelle and Sidoti, Olivia},
	month = dec,
	year = {2024},
}

@article{tidy_tiktok_2020,
	title = {{TikTok}: {The} story of a social media giant},
	shorttitle = {{TikTok}},
	url = {https://www.bbc.com/news/technology-53640724},
	language = {en-GB},
	urldate = {2025-08-14},
	journal = {BBC News},
	author = {Tidy, Joe and Smith Galer, Sophia},
	month = aug,
	year = {2020},
}

@inproceedings{chandio2024audit,
  title={How audit methods impact our understanding of youtube’s recommendation systems},
  author={Chandio, Sarmad and Dar, Muhammad Daniyal Pirwani and Nithyanand, Rishab},
  booktitle={Proceedings of the International AAAI Conference on Web and Social Media},
  volume={18},
  pages={241--253},
  year={2024}
}

@inproceedings{sharevski2025lowvision,
  title={``I Don’t Think TikTok Really Cares About the Truth'': Experiences of Users Who Are Low Vision or Blind with Misinformation on TikTok},
  author={Sharevski, Filip and Zeidieh, Aida},
  booktitle={Proceedings of the International AAAI Conference on Web and Social Media},
  volume={19},
  number={1},
  pages={1786--1797},
  year={2025},
  doi={10.1609/icwsm.v19i1.35901}
}

@inproceedings{galdeman2025climate,
  title={Mapping the Climate Change Landscape on TikTok},
  author={Galdeman, Alex and Aiello, Luca Maria},
  booktitle={Proceedings of the International AAAI Conference on Web and Social Media},
  volume={19},
  number={1},
  pages={2614--2621},
  year={2025},
  doi={10.1609/icwsm.v19i1.35962}
}

@inproceedings{ungless2025censorship,
  title={Experiences of Censorship on TikTok Across Marginalised Identities},
  author={Ungless, Emma L. and Markl, Nina and Ross, Bj{\"o}rn},
  booktitle={Proceedings of the International AAAI Conference on Web and Social Media},
  volume={19},
  number={1},
  pages={1952--1965},
  year={2025},
  doi={10.1609/icwsm.v19i1.35912}
}

@inproceedings{yang2025addiction,
  title={Studying Behavioral Addiction by Combining Surveys and Digital Traces: A Case Study of TikTok},
  author={Yang, Cheng and Mousavi, Seyedmehdi and Dash, Abhishek and Gummadi, Krishna P. and Weber, Ingmar},
  booktitle={Proceedings of the International AAAI Conference on Web and Social Media},
  volume={19},
  number={1},
  pages={2106--2123},
  year={2025},
  doi={10.1609/icwsm.v19i1.35922}
}

@incollection{newell1973you,
  title={You can't play 20 questions with nature and win: Projective comments on the papers of this symposium},
  author={Newell, Allen},
  booktitle={Machine intelligence},
  pages={121--146},
  year={1973},
  publisher={Routledge}
}

@article{watts2017should,
  title={Should social science be more solution-oriented?},
  author={Watts, Duncan J},
  journal={Nature Human Behaviour},
  volume={1},
  number={1},
  pages={0015},
  year={2017},
  publisher={Nature Publishing Group UK London}
}

@article{rejeb_mapping_2024,
	title = {Mapping the scholarly landscape of {TikTok} ({Douyin}): {A} bibliometric exploration of research topics and trends},
	volume = {4},
	issn = {26669544},
	shorttitle = {Mapping the scholarly landscape of {TikTok} ({Douyin})},
	url = {https://linkinghub.elsevier.com/retrieve/pii/S2666954424000036},
	doi = {10.1016/j.digbus.2024.100075},
	language = {en},
	number = {1},
	urldate = {2025-08-27},
	journal = {Digital Business},
	author = {Rejeb, Abderahman and Rejeb, Karim and Appolloni, Andrea and Treiblmaier, Horst and Iranmanesh, Mohammad},
	month = jun,
	year = {2024},
	pages = {100075},
}

@article{kanthawala_its_2022,
	title = {It’s the {Methodology} {For} {Me}: {A} {Systematic} {Review} of {Early} {Approaches} to {Studying} {TikTok}},
	shorttitle = {It’s the {Methodology} {For} {Me}},
	url = {https://aisel.aisnet.org/hicss-55/dsm/digital_methods/4},
	journal = {Hawaii International Conference on System Sciences 2022 (HICSS-55)},
	author = {Kanthawala, Shaheen and Cotter, Kelley and Foyle, Kali and Decook, J. R.},
	month = jan,
	year = {2022},
}

@misc{tiktok_tiktok_2026,
	title = {{TikTok} {Research} {Tools} {Terms} of {Service} {\textbar} {TikTok}},
	url = {https://www.tiktok.com/legal/page/global/terms-of-service-research-api/en},
	urldate = {2026-05-08},
	journal = {https://www.tiktok.com},
	author = {TikTok},
	month = jan,
	year = {2026},
}

@article{wedel_platform_2026,
	title = {The platform matters: cross-platform differences in data donation willingness, behavior, and bias},
	volume = {20},
	issn = {1931-2458},
	shorttitle = {The platform matters},
	url = {https://doi.org/10.1080/19312458.2025.2605946},
	doi = {10.1080/19312458.2025.2605946},
	number = {1},
	urldate = {2026-05-09},
	journal = {Communication Methods and Measures},
	publisher = {Routledge},
	author = {Wedel, Lion and Ohme, Jakob and Mayer, Anna-Theresa and Gaisbauer, Felix and Fan, Yangliu},
	month = jan,
	year = {2026},
	note = {\_eprint: https://doi.org/10.1080/19312458.2025.2605946},
	pages = {53--77},
}

@article{Freelon2025,
  author       = {Freelon, D. and Monzer, C. and Jeon, G. and Moy, C. and Williams, N.},
  title        = {The Post-API Age of Social Media Data Access: Past, Present, and Future},
  journal      = {The ANNALS of the American Academy of Political and Social Science},
  year         = {2025},
  volume       = {715},
  number       = {1},
  pages        = {16--37},
  doi          = {10.1177/00027162251372557},
  url          = {https://doi.org/10.1177/00027162251372557}
}

@misc{MetaBrightData_OrderShortenTime2023,
  key          = {Meta Platforms v. Bright Data},
  title        = {{Meta Platforms, Inc. v. Bright Data Ltd.}: Order on Motion to Shorten Time},
  howpublished = {\url{https://www.govinfo.gov/content/pkg/USCOURTS-cand-3_23-cv-00077/pdf/USCOURTS-cand-3_23-cv-00077-0.pdf}},
  year         = {2023},
  note         = {U.S. District Court for the Northern District of California, Case No. 23-cv-00077-AGT},
  url          = {https://www.govinfo.gov/content/pkg/USCOURTS-cand-3_23-cv-00077/pdf/USCOURTS-cand-3_23-cv-00077-0.pdf}
}

@misc{SandvigSummaryJudgment2020,
  key          = {Sandvig v. Barr},
  title        = {{Sandvig v. Barr}: Summary Judgment Opinion},
  howpublished = {\url{https://assets.aclu.org/live/uploads/legal-documents/Sandvig_Summary_Judgment_Opinion.pdf}},
  year         = {2020},
  note         = {U.S. District Court for the District of Columbia, No. 1:16-cv-01368},
  url          = {https://assets.aclu.org/live/uploads/legal-documents/Sandvig_Summary_Judgment_Opinion.pdf}
}

@article{Freelon2018,
  author       = {Freelon, Deen},
  title        = {Computational Research in the Post-{API} Age},
  journal      = {Political Communication},
  year         = {2018},
  volume       = {35},
  number       = {4},
  pages        = {665--668},
  publisher    = {Routledge},
  doi          = {10.1080/10584609.2018.1477506}
}

@article{suarez-alvarez_communicational_2022,
	title = {Communicational analysis about building of unequal relationships on social networks. {Case} \#sugardaddy in {TikTok}},
	volume = {27},
	shorttitle = {Análisis comunicacional de la construcción de las relaciones no igualitarias en las redes sociales. {Caso} \#sugardaddy en {TikTok}},
	url = {https://www.scopus.com/inward/record.uri?eid=2-s2.0-85143142192&doi=10.5209%2fhics.84389&partnerID=40&md5=bf381313e015883341273f940fad1c2f},
	doi = {10.5209/hics.84389},
	number = {2},
	journal = {Historia y Comunicacion Social},
	author = {Suárez-Álvarez, R.},
	year = {2022},
	pages = {401--413},
}

@article{meza_idols_2023,
	title = {Idols of {Promotion} and {Authenticity} on {TikTok}},
	volume = {11},
	url = {https://www.scopus.com/inward/record.uri?eid=2-s2.0-85178124538&doi=10.17645%2fmac.v11i4.7123&partnerID=40&md5=d030223da9ce5f3fb54c53aabf2a5075},
	doi = {10.17645/mac.v11i4.7123},
	number = {4},
	journal = {Media and Communication},
	author = {Meza, R.M. and Mogoș, A.-A. and Prundaru, G.},
	year = {2023},
	pages = {187--202},
}

@article{sun_vaping_2023,
	title = {Vaping on {TikTok}: a systematic thematic analysis},
	volume = {32},
	url = {https://www.scopus.com/inward/record.uri?eid=2-s2.0-85111433640&doi=10.1136%2ftobaccocontrol-2021-056619&partnerID=40&md5=e5cdf2eb05da209f614bd27a4b765d21},
	doi = {10.1136/tobaccocontrol-2021-056619},
	number = {2},
	journal = {Tobacco Control},
	author = {Sun, T. and Lim, C.C.W. and Chung, J. and Cheng, B. and Davidson, L. and Tisdale, C. and Leung, J. and Gartner, C.E. and Connor, J. and Hall, W.D. and Chan, G.C.K.},
	year = {2023},
	pages = {251--254},
}

@article{infante_pineda_journalists_2025,
	title = {The journalist’s profile and their digital competence in social media: the case study of Ángel {Martín} on {TikTok}},
	volume = {2025-January-June},
	shorttitle = {Perfil periodístico y competencias digitales en redes sociales: estudio de caso de Ángel {Martín} en {TikTok}},
	url = {https://www.scopus.com/inward/record.uri?eid=2-s2.0-85215118533&doi=10.31921%2fdoxacom.n40a2723&partnerID=40&md5=876f077877c60d81db2109bf1c83680b},
	doi = {10.31921/doxacom.n40a2723},
	number = {40},
	journal = {Doxa Comunicacion},
	author = {Infante Pineda, S. and García Vega, A. and Martínez Borda, R. and Barrajón Lara, I.},
	year = {2025},
	pages = {435--464},
}

@article{izquierdo-condoy_assessing_2025,
	title = {Assessing the educational impact and quality of medical microvideos on {TikTok}: the case of {Latin} {America}},
	volume = {30},
	url = {https://www.scopus.com/inward/record.uri?eid=2-s2.0-105000000535&doi=10.1080%2f10872981.2025.2474129&partnerID=40&md5=275253ebcbe4e53749791b13a9c71301},
	doi = {10.1080/10872981.2025.2474129},
	number = {1},
	journal = {Medical Education Online},
	author = {Izquierdo-Condoy, J.S. and Arias-Intriago, M. and Mosquera-Quiñónez, M. and Melgar Muñoz, F.P. and Jiménez-Ascanio, M. and Loaiza-Guevara, V. and Ortiz-Prado, E.},
	year = {2025},
}

@article{santaolalla-rueda_potaxies_2024,
	title = {Potaxies and {Fifes}: {The} {Formation} of {New} {Subcultures} on {TikTok}},
	volume = {14},
	url = {https://www.scopus.com/inward/record.uri?eid=2-s2.0-85213479856&doi=10.3390%2fsoc14120265&partnerID=40&md5=5212cf77186778ff4d9d6e79ee2649e1},
	doi = {10.3390/soc14120265},
	number = {12},
	journal = {Societies},
	author = {Santaolalla-Rueda, P. and Fernández-Muñoz, C.},
	year = {2024},
}

@article{lee_identification_2025,
	title = {Identification and {Classification} of {Images} in e-{Cigarette}-{Related} {Content} on {TikTok}: {Unsupervised} {Machine} {Learning} {Image} {Clustering} {Approach}},
	volume = {60},
	url = {https://www.scopus.com/inward/record.uri?eid=2-s2.0-105001799808&doi=10.1080%2f10826084.2024.2447415&partnerID=40&md5=41ca44176b28b14f432008f082c54094},
	doi = {10.1080/10826084.2024.2447415},
	number = {5},
	journal = {Substance Use and Misuse},
	author = {Lee, J. and Murthy, D. and Ouellette, R. and Anand, T. and Kong, G.},
	year = {2025},
	pages = {677--683},
}

@article{cartes-barroso_attracting_2025,
	title = {Attracting the {Vote} on {TikTok}: {Far}-{Right} {Parties}’ {Emotional} {Communication} {Strategies} in the 2024 {European} {Elections}},
	volume = {6},
	url = {https://www.scopus.com/inward/record.uri?eid=2-s2.0-105001125903&doi=10.3390%2fjournalmedia6010033&partnerID=40&md5=42eae42f8ccab3c555dae8bd31c45159},
	doi = {10.3390/journalmedia6010033},
	number = {1},
	journal = {Journalism and Media},
	author = {Cartes-Barroso, M.J. and García-Estévez, N. and Méndez-Muros, S.},
	year = {2025},
}

@article{londono-moreno_peruvian_2025,
	title = {Peruvian minors and their exposure to beauty and makeup content on social networks: {A} content analysis on {YouTube}, {Instagram}, and {TikTok}},
	volume = {16},
	shorttitle = {Menores peruanos y su exposición al contenido de belleza y maquillaje en redes sociales: análisis del contenido en {YouTube}, {Instagram} y {TikTok}},
	url = {https://www.scopus.com/inward/record.uri?eid=2-s2.0-85214134763&doi=10.14198%2fMEDCOM.27875&partnerID=40&md5=847ae17904959f570bee94d9d9637511},
	doi = {10.14198/MEDCOM.27875},
	number = {1},
	journal = {Revista Mediterranea de Comunicacion},
	author = {Londoño-Moreno, M. and Atarama-Rojas, T. and Anastacio-Coello, L.},
	year = {2025},
}

@article{caro-castano_self-presentation_2025,
	title = {Self-{Presentation} and {Engagement} of {Left}-{Wing} {Politicians} on {Tiktok} and {Instagram}: {A} {Comparative} {Study} of {Female} {MPs} in {Argentina} and {Spain}},
	volume = {44},
	shorttitle = {Autoapresentação e engajamento de políticos de esquerda no {Tiktok} e no {Instagram}: um estudo comparativo de deputadas da {Argentina} e da {Espanha}},
	url = {https://www.scopus.com/inward/record.uri?eid=2-s2.0-105007316367&doi=10.11144%2fJaveriana.syp44.aepi&partnerID=40&md5=bfcf135533c4b0fe1d139a26409761b8},
	doi = {10.11144/Javeriana.syp44.aepi},
	journal = {Signo y Pensamiento},
	author = {Caro-Castaño, L. and Slimovich, A.},
	year = {2025},
}

@article{steinke_women_2024,
	title = {Women in {STEM} on {TikTok}: {Advancing} {Visibility} and {Voice} {Through} {STEM} {Identity} {Expression}},
	volume = {10},
	url = {https://www.scopus.com/inward/record.uri?eid=2-s2.0-85201294711&doi=10.1177%2f20563051241274675&partnerID=40&md5=db99011018fd077abed80779f1e2efc2},
	doi = {10.1177/20563051241274675},
	number = {3},
	journal = {Social Media and Society},
	author = {Steinke, J. and Gilbert, C. and Coletti, A. and Levin, S.H. and Suk, J. and Oeldorf-Hirsch, A.},
	year = {2024},
}

@article{scharlach_how_2024,
	title = {How to {Spark} {Joy}: {Strategies} of {Depoliticization} in {Platform}’s {Corporate} {Social} {Initiatives}},
	volume = {10},
	url = {https://www.scopus.com/inward/record.uri?eid=2-s2.0-85203428379&doi=10.1177%2f20563051241277601&partnerID=40&md5=e46f7eafca84624ceffa56c7c7b8194c},
	doi = {10.1177/20563051241277601},
	number = {3},
	journal = {Social Media and Society},
	author = {Scharlach, R.},
	year = {2024},
}

@article{luik_media_2025,
	title = {Media logic and educational micro-content: {Presentational} themes and approaches on {TikTok}},
	volume = {28},
	url = {https://www.scopus.com/inward/record.uri?eid=2-s2.0-85216450795&doi=10.1080%2f10714421.2025.2452086&partnerID=40&md5=0c25a505a82cfb04b5e7a33734228f67},
	doi = {10.1080/10714421.2025.2452086},
	number = {2},
	journal = {Communication Review},
	author = {Luik, J. and Setiawan, D. and Sitindjak, R.H.I.},
	year = {2025},
	pages = {170--196},
}

@article{dominguez-garcia_emotion_2025,
	title = {From {Emotion} to {Virality}: {The} {Use} of {Social} {Media} in the {Populism} of {Chega} and {VOX}},
	volume = {14},
	url = {https://www.scopus.com/inward/record.uri?eid=2-s2.0-105006417826&doi=10.3390%2fsocsci14050255&partnerID=40&md5=86c584b0cedb2b797834b5ab1fc508a7},
	doi = {10.3390/socsci14050255},
	number = {5},
	journal = {Social Sciences},
	author = {Domínguez-García, R. and Baptista, J.P. and Pérez-Curiel, C. and Fonseca, D.E.M.D.},
	year = {2025},
}

@article{ji_revealing_2025,
	title = {Revealing public attitudes toward ‘substituting plastic with bamboo’ in {China}: {Sentiment} and topic analyses using social media data},
	volume = {176},
	url = {https://www.scopus.com/inward/record.uri?eid=2-s2.0-105004547372&doi=10.1016%2fj.forpol.2025.103508&partnerID=40&md5=aa692cc2b1ac16d90d610c9de17852a1},
	doi = {10.1016/j.forpol.2025.103508},
	journal = {Forest Policy and Economics},
	author = {Ji, J. and Xu, X. and Tam, V.W.Y. and Zhang, Y.},
	year = {2025},
}

@article{jerin_mental_2024,
	title = {Mental health messages on {TikTok}: {Analysing} the use of emotional appeals in health-related \#{EduTok} videos},
	volume = {83},
	url = {https://www.scopus.com/inward/record.uri?eid=2-s2.0-85188325131&doi=10.1177%2f00178969241235528&partnerID=40&md5=cc7ad5cdf339e851d02f52c59d96cd19},
	doi = {10.1177/00178969241235528},
	number = {4},
	journal = {Health Education Journal},
	author = {Jerin, S.I. and O’Donnell, N. and Mu, D.},
	year = {2024},
	pages = {395--408},
}

@article{zhang_impact_2025,
	title = {The impact of platform affordance on the representation of iconic architecture: a case study of {Markthal} across visual social media},
	volume = {19},
	url = {https://www.scopus.com/inward/record.uri?eid=2-s2.0-85214827799&doi=10.1108%2fARCH-05-2024-0198&partnerID=40&md5=238f6ec073a67ddae7a410c647180c26},
	doi = {10.1108/ARCH-05-2024-0198},
	number = {2},
	journal = {Archnet-IJAR: International Journal of Architectural Research},
	author = {Zhang, S. and Zhang, G.},
	year = {2025},
	pages = {390--410},
}

@article{pearson_beyond_2025,
	title = {Beyond the margin of error: a systematic and replicable audit of the {TikTok} research {API}},
	volume = {28},
	issn = {1369-118X},
	shorttitle = {Beyond the margin of error},
	url = {https://doi.org/10.1080/1369118X.2024.2420032},
	doi = {10.1080/1369118X.2024.2420032},
	number = {3},
	urldate = {2025-05-30},
	journal = {Information, Communication \& Society},
	author = {Pearson, George D.H. and , Nathan A., Silver and , Jessica Y., Robinson and , Mona, Azadi and , Barbara A., Schillo and and Kreslake, Jennifer M.},
	month = feb,
	year = {2025},
	note = {Publisher: Routledge
\_eprint: https://doi.org/10.1080/1369118X.2024.2420032},
	pages = {452--470},
}

@inproceedings{corso_what_2024,
	address = {Stuttgart Germany},
	title = {What we can learn from {TikTok} through its {Research} {API}},
	isbn = {979-8-4007-0453-6},
	url = {https://dl.acm.org/doi/10.1145/3630744.3663611},
	doi = {10.1145/3630744.3663611},
	language = {en},
	urldate = {2025-05-30},
	booktitle = {Companion {Proceedings} of the 16th {ACM} {Web} {Science} {Conference}},
	publisher = {ACM},
	author = {Corso, Francesco and Pierri, Francesco and De Francisci Morales, Gianmarco},
	month = may,
	year = {2024},
	pages = {110--114},
}

@online{freelon_pyktok_nodate,
  title        = {Pyktok: TikTok Data Collection Tools},
  author       = {Freelon, Deen},
  year         = {n.d.},
  url          = {https://github.com/dfreelon/pyktok},
  urldate      = {2025-10-01},
  organization = {GitHub},
  note         = {Open-source TikTok data collection tools},
}

@online{tiktok_developers_research_api,
  author = {{TikTok for Developers}},
  title = {{Research API Specs - Query Video Comments}},
  year = {2025},
  url = {https://developers.tiktok.com/doc/research-api-specs-query-video-comments},
  urldate = {2025-10-02}
}

@online{tiktok_api_v2_error_handling,
  title = {{TikTok API v2 Error Handling}},
  author = {{TikTok for Developers}},
  year = {2025},
  url = {https://developers.tiktok.com/doc/tiktok-api-v2-error-handling},
  urldate = {2025-10-02}
}

@online{tiktok_research_api_wrapper,
  author = {{TikTok for Developers}},
  title = {{Introducing TikTok Research API Wrappers on GitHub}},
  year = {2025},
  url = {https://developers.tiktok.com/blog/research-api-wrappers-on-github-announcement},
  urldate = {2025-10-02}
}

@inproceedings{kaplan2026foryou,
  title={When `For You' Isn't for You: Measuring User Agency in TikTok's Algorithmic Feed},
  author={Kaplan, Levi and Patel, Devin and Gerzon, Nicole and Mislove, Alan and Sapiezynski, Piotr},
  booktitle={Proceedings of the International AAAI Conference on Web and Social Media},
  volume={20},
  number={1},
  pages={1175--1189},
  year={2026},
  doi={10.1609/icwsm.v20i1.42688}
}

@inproceedings{bickham2026tied,
  title={Tied In on TikTok: Tie Strength and Emotional Dynamics in Algorithmic Communities},
  author={Bickham, Charles and Chu, Minh Duc and Yuan, Arianna and Lookingbill, Valerie and Mohammadi, Ehsan and Murray, Stuart and Lerman, Kristina and Ferrara, Emilio},
  booktitle={Proceedings of the International AAAI Conference on Web and Social Media},
  volume={20},
  number={1},
  pages={262--275},
  year={2026},
  doi={10.1609/icwsm.v20i1.42637}
}

@inproceedings{xu2026anti,
  title={Anti-Establishment Sentiment on TikTok: Implications for Understanding Influence(rs) and Expertise on Social Media},
  author={Xu, Tianliang and Hasell, Ariel and Tomkins, Sabina},
  booktitle={Proceedings of the International AAAI Conference on Web and Social Media},
  volume={20},
  number={1},
  pages={2538--2553},
  year={2026},
  doi={10.1609/icwsm.v20i1.42765}
}

@inproceedings{luceri2026coordinated,
  title={Coordinated Inauthentic Behavior on TikTok: Challenges and Opportunities for Detection in a Video-First Ecosystem},
  author={Luceri, Luca and Salkar, Tanishq Vijay and Balasubramanian, Ashwin and Pinto, Gabriela and Sun, Chenning and Ferrara, Emilio},
  booktitle={Proceedings of the International AAAI Conference on Web and Social Media},
  volume={20},
  number={1},
  pages={1533--1550},
  year={2026},
  doi={10.1609/icwsm.v20i1.42711}
}

\section{Ethics Checklist}

\begin{enumerate}

\item For most authors...
\begin{enumerate}
    \item  Would answering this research question advance science without violating social contracts, such as violating privacy norms, perpetuating unfair profiling, exacerbating the socio-economic divide, or implying disrespect to societies or cultures?
    \answerYes{Yes}
  \item Do your main claims in the abstract and introduction accurately reflect the paper's contributions and scope?
    \answerYes{Yes}
   \item Do you clarify how the proposed methodological approach is appropriate for the claims made? 
    \answerYes{Yes}
   \item Do you clarify what are possible artifacts in the data used, given population-specific distributions?
    \answerYes{Yes}
  \item Did you describe the limitations of your work?
    \answerYes{Yes}
  \item Did you discuss any potential negative societal impacts of your work?
    \answerYes{Yes}
      \item Did you discuss any potential misuse of your work?
    \answerYes{Yes}
    \item Did you describe steps taken to prevent or mitigate potential negative outcomes of the research, such as data and model documentation, data anonymization, responsible release, access control, and the reproducibility of findings?
    \answerYes{Yes}
  \item Have you read the ethics review guidelines and ensured that your paper conforms to them?
    \answerYes{Yes}
\end{enumerate}

\item Additionally, if you are using existing assets (e.g., code, data, models) or curating/releasing new assets, \textbf{without compromising anonymity}...
\begin{enumerate}
  \item If your work uses existing assets, did you cite the creators?
    \answerYes{Yes}
  \item Did you mention the license of the assets?
    \answerYes{Yes}
  \item Did you include any new assets in the supplemental material or as a URL?
    \answerNo{No, because this study does not release new code, data, or software.}
  \item Did you discuss whether and how consent was obtained from people whose data you're using/curating?
    \answerYes{Yes, and the study relies exclusively on publicly accessible content collected through official or publicly documented interfaces.}
  \item Did you discuss whether the data you are using/curating contains personally identifiable information or offensive content?
    \answerYes{Yes, and while the data may contain user-generated identifiers or content, we do not attempt to identify individuals and conduct analyses at an aggregate level.}
\item If you are curating or releasing new datasets, did you discuss how you intend to make your datasets FAIR?
\answerNA{NA}

\item If you are curating or releasing new datasets, did you create a Datasheet for the Dataset? 
\answerNA{NA}

\end{enumerate}
\end{enumerate}

\newpage
\appendix
\section{Diagnostic Test of API(R) Pagination}

Table~\ref{tab:api_random_pagination_replication} reports an April 24 diagnostic test using March 1--31, 2026 as the query window. The test reproduces the pagination pattern observed during the original collection. With \texttt{fetch\_all\_pages=True}, API(D) returned \texttt{has\_more=True} for every matched hashtag and keyword query, whereas API(R) returned \texttt{has\_more=False} for every matched query. Thus, API(R) did not provide the continuation fields needed for pagination, while API(D) did.

\begin{table}[H]
\centering
\footnotesize
\setlength{\tabcolsep}{4pt}
\begin{tabular}{llccrr}
\toprule
Search type & Target &
\multicolumn{2}{c}{\texttt{has\_more}} &
\multicolumn{2}{c}{Unique IDs} \\
\cmidrule(lr){3-4} \cmidrule(lr){5-6}
& & API(D) & API(R) & API(D) & API(R) \\
\midrule
hashtag & fyp     & True & False & 1034 & 178 \\
hashtag & news    & True & False & 1061 & 195 \\
hashtag & trump   & True & False & 1081 & 192 \\
hashtag & vaccine & True & False & 1078 & 193 \\
keyword & fyp     & True & False & 1024 & 180 \\
keyword & news    & True & False & 1037 & 180 \\
keyword & trump   & True & False & 1057 & 187 \\
keyword & vaccine & True & False & 1057 & 190 \\
\bottomrule
\end{tabular}
\caption{Diagnostic test of API(D) and API(R) pagination behavior under \texttt{fetch\_all\_pages=True} on April 24, 2026.}
\label{tab:api_random_pagination_replication}
\end{table}

\section{Additional Engagement Distribution Figures}

Figures~\ref{fig:appendix_comment_ridgeline}--\ref{fig:appendix_view_ridgeline} provide metric-specific ridgeline distributions supplementing Figure~\ref{fig:hashtag_keyword_engagement_dist}. In each figure, columns distinguish API(D), API(R), and Apify, while rows distinguish hashtag and keyword endpoints. Distributions are shown separately for fyp, news, trump, and vaccine. The x-axis is log10 scaled, and vertical lines indicate medians.

\begin{figure}[H]
    \centering
    \includegraphics[width=1\columnwidth]{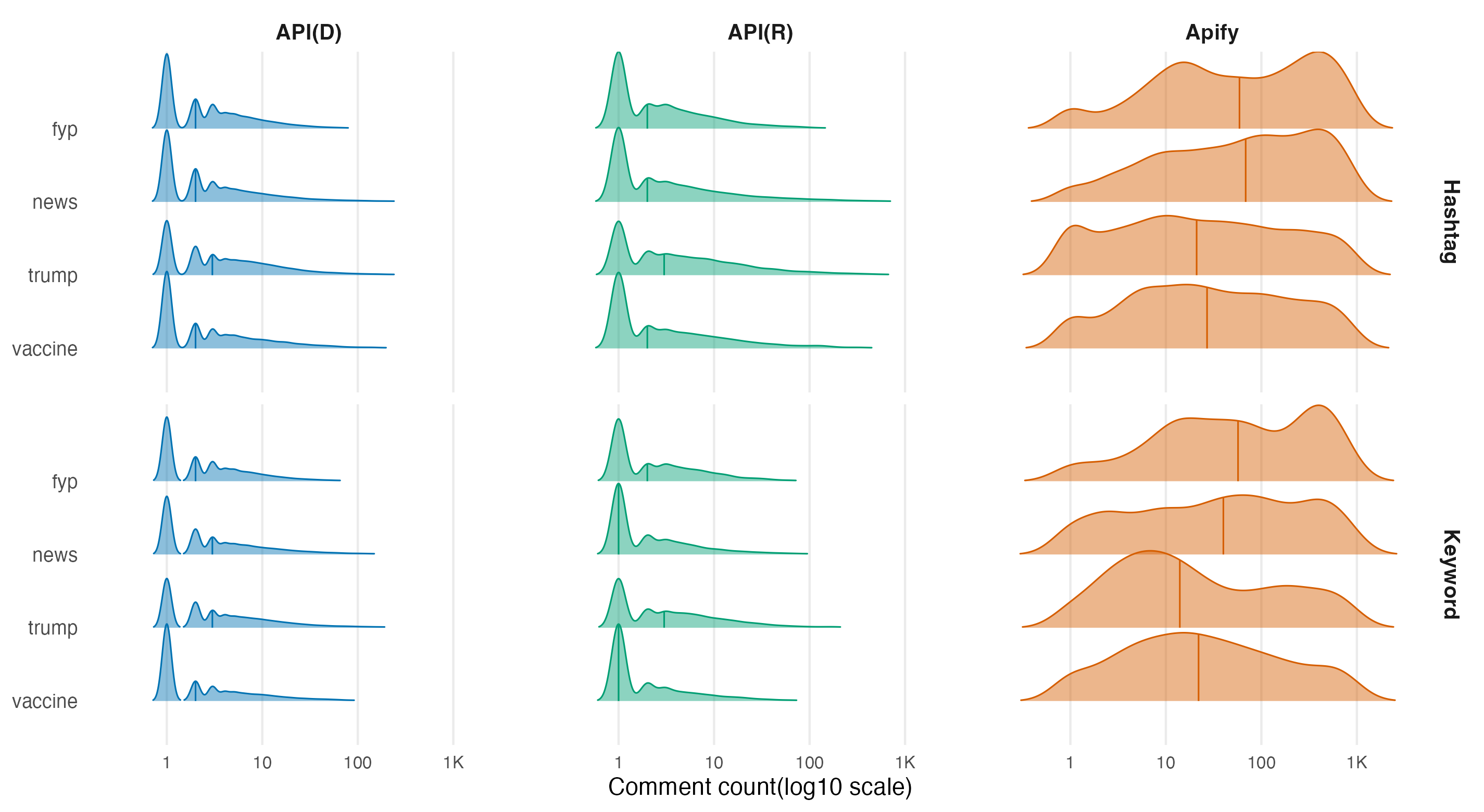}
    \caption{Comment count distributions. API(D) and API(R) are concentrated near the lower end of the scale, while Apify returns videos with higher comment counts across most targets.}
    \label{fig:appendix_comment_ridgeline}
\end{figure}

\begin{figure}[H]
    \centering
    \includegraphics[width=1\columnwidth]{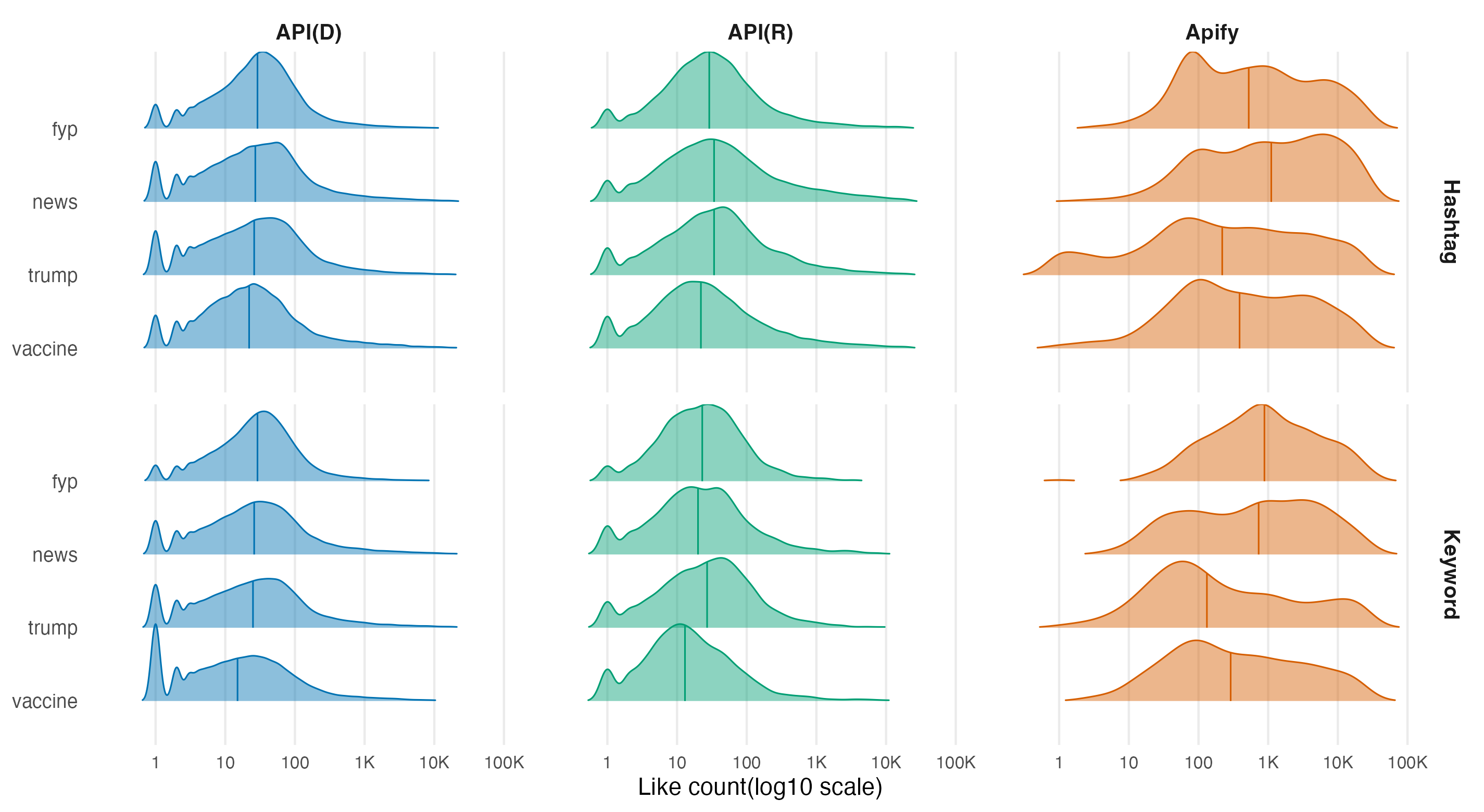}
    \caption{Like count distributions. API(D) and API(R) exhibit broadly similar distributions, whereas Apify is shifted toward higher like counts.}
    \label{fig:appendix_like_ridgeline}
\end{figure}

\begin{figure}[H]
    \centering
    \includegraphics[width=1\columnwidth]{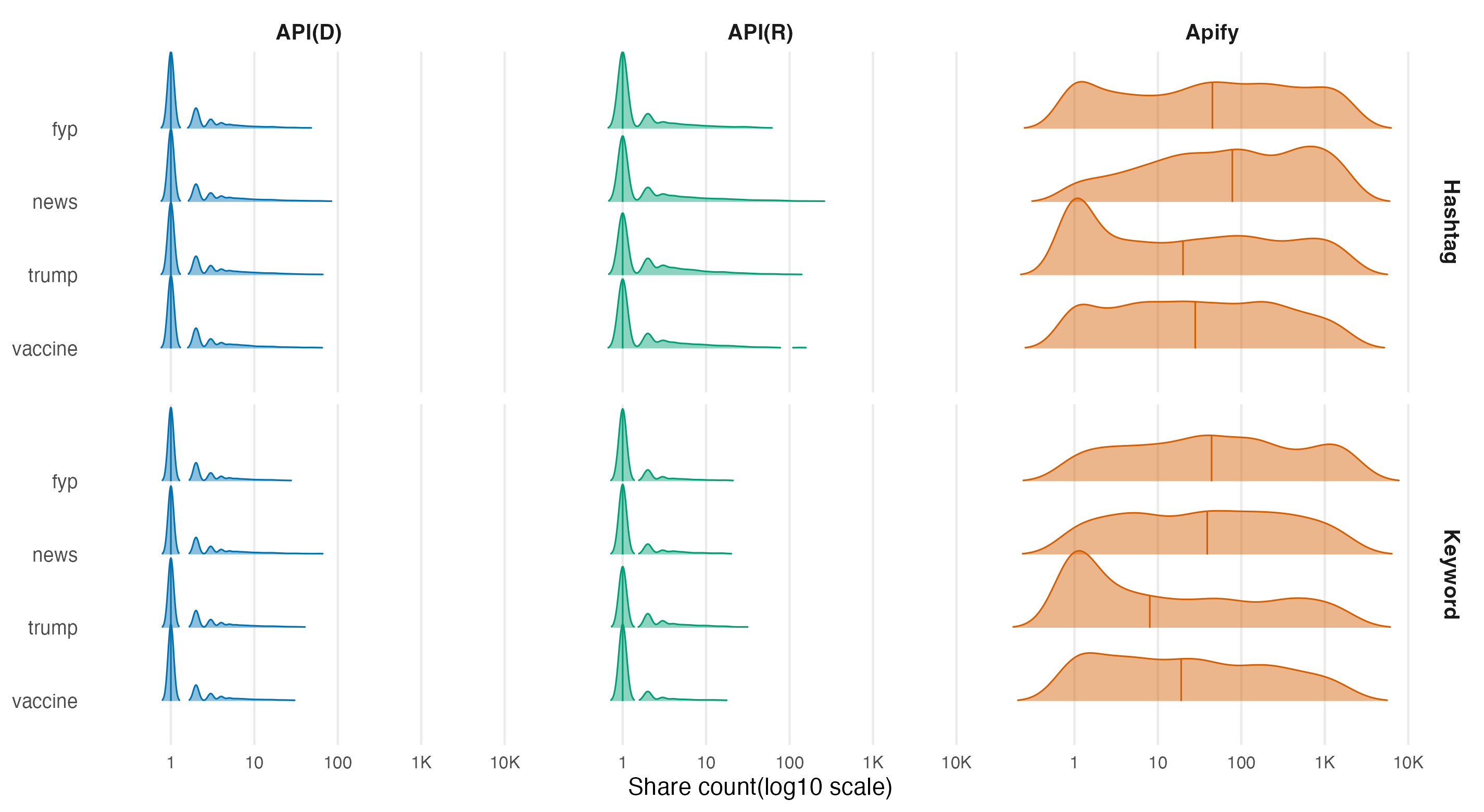}
    \caption{Share count distributions. Both API modes are concentrated near low values, while Apify shows higher medians and broader high engagement tails.}
    \label{fig:appendix_share_ridgeline}
\end{figure}

\begin{figure}[H]
    \centering
    \includegraphics[width=1\columnwidth]{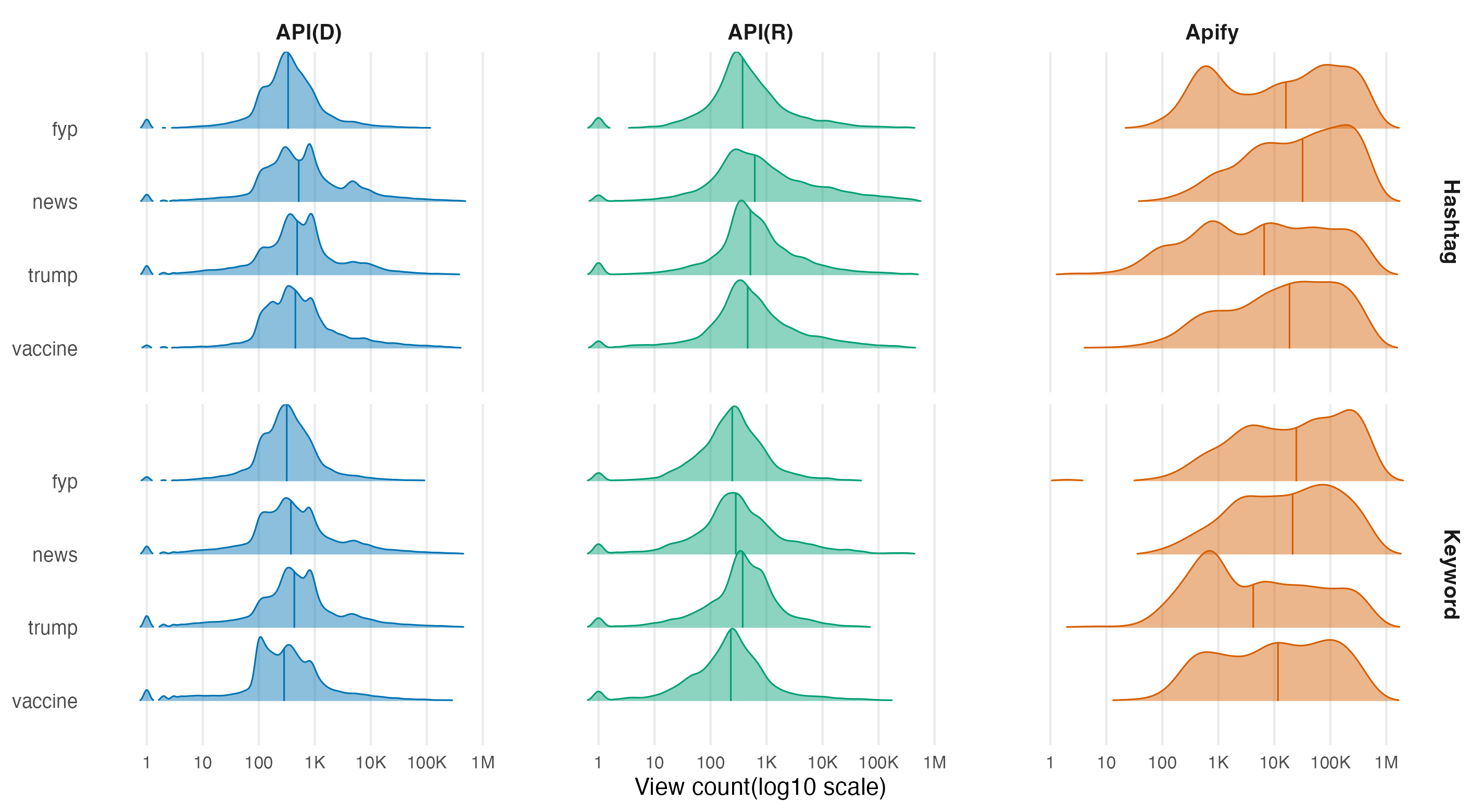}
    \caption{View count distributions. Apify returns videos with higher view count distributions than API(D) and API(R) across both endpoints.}
    \label{fig:appendix_view_ridgeline}
\end{figure}

\end{document}